\documentclass{jpp}
\usepackage{graphicx}
\usepackage{xcolor}
\usepackage{bm}
\newcommand{\partder}[2]{\frac{\partial #1}{\partial #2}}

\usepackage[utf8]{inputenc}
\usepackage[T1]{fontenc}
\usepackage{amsmath}
\usepackage{subcaption}

\title{Fast ion transport in quasisymmetric equilibria in the presence of a resonant Alfv\'{e}nic perturbation}

\author{Elizabeth J. Paul 
  \corresp{\email{ejp2170@columbia.edu}}\aff{1}, Harry E. Mynick\aff{2}, Amitava Bhattacharjee\aff{2}}

\affiliation{\aff{1}Columbia University, New York, NY 10027, USA \\ \aff{2}Princeton Plasma Physics Laboratory, Princeton, NJ 08540, USA}

\begin{document}

\maketitle

\begin{abstract}
Significant progress has been made in designing magnetic fields that provide excellent confinement of the guiding enter trajectories of alpha particles using quasisymmetry (QS). Given the reduction in this transport channel, we assess the impact of resonant Alfv\'{e}n eigenmodes (AEs) on the guiding center motion. The AE amplitudes are chosen to be consistent with experimental measurements and large-scale simulations. We evaluate the drift resonance condition, phase-space island width, and island overlap criterion for quasisymmetric configurations. Kinetic Poincar\'{e} plots elucidate features of the transport, including stiff transport above a critical perturbation amplitude. Our analysis highlights key departures from the AE-driven transport in tokamaks, such as the avoidance of phase-space island overlap in quasihelical configurations and the enhanced transport due to wide phase-space islands in low magnetic shear configurations. In configurations that are closer to QS, with QS deviations $\delta B/B_0 \lesssim 10^{-3}$, the transport is primarily driven by the AE, while configurations that are further from QS, $\delta B/B_0 \sim 10^{-2}$, experience significant transport due to the QS-breaking fields in addition to the AE. 
\end{abstract}

\section{Introduction}

Energetic particles have historically been challenging to confine in stellarator configurations due to the possibility of unconfined orbits and resonances exposed at low collisionality. These difficulties must be overcome to develop a stellarator reactor concept, as excessive alpha losses before thermalization can impact power balance and impart damage to plasma-facing components. In recent years, significant progress has been made in designing stellarator magnetic fields that can confine the orbits of fusion-born alpha particles without perturbations \citep{2019Bader,2022Landreman,2022Landremanb}. However, reducing the guiding center orbit loss mechanisms may make mode-particle interactions relatively significant. 

The interaction of Alfv\'{e}n eigenmodes (AEs) with energetic particles has been shown to drive substantial flattening of the fast-ion profile in tokamak experiments \citep{2008Heidbrink}. Alv\'{e}nic activity has also been observed on several stellarator configurations, including HSX \citep{2009Deng}, CHS \citep{2002Kakechi}, LHD \citep{2011Toi}, W7-AS \citep{1994Weller}, TJ-II \citep{2014Melnikov}, W7-X \citep{2020Rahbarnia}, and Heliotron-J \citep{2007Yamamoto}.
Alfv\'{e}nic instabilities have been thought to be potentially benign in a stellarator reactor due to their ability to operate at high density \citep{2012Helander}. 
However, fast-ion-driven modes may still be destabilized at high density: LHD modeling indicates that Alfv\'{e}nic activity remains present even at fast ion beta of $\approx 0.05\%$ \citep{2017Varela}. 
For comparison, using the profiles from the ARIES-CS stellarator reactor study with density $\approx 5 \times 10^{20}$ m$^{-3}$ \citep{2008Ku}, the fast ion beta is $\approx 0.2\%$. It remains to be seen to what extent Alfv\'{e}nic activity can be controlled in a stellarator reactor by manipulating the density profile or optimizing the magnetic field. 

Significant recent work has focused on energetic particle physics in quasisymmetric configurations based on properties of guiding center orbits \citep{2021Bader,2022Leviness,2022Paul}, but relatively little has been studied with respect to AE-driven transport. 
AE stability of the quasiaxisymmetric CFQS configuration has been evaluated using the linear gyrofluid FAR3D code \citep{2021Varela}. 
However, a systematic study of the AE-driven transport in quasisymmetric configurations has not yet been performed.
For equilibria sufficiently close to quasisymmetry (QS), phenomena previously observed on tokamaks are anticipated. 
Monte Carlo simulations indicate that a resonant perturbation induces some rapid convective transport due to phase-space islands near the boundaries. Island overlap occurs for larger perturbation amplitudes, $\delta B^r/B_0 \sim 10^{-3}$ where $\delta B^r$ is the radial perturbed magnetic field and $B_0$ is the equilibrium field strength, causing diffusive losses \citep{1992Sigmar,1992Hsu}. Because of the coupling of a single AE to the poloidal variation of the magnetic drifts, sideband resonances arise and can lead to island overlap even in the presence of a single AE \citep{1993Mynicka,1993Mynickb,1992Hsu}. In perfect symmetry with the addition of a single perturbation harmonic, kinetic Poincar\'{e} plots \citep{2011White,2012White} can be employed to observe the formation of phase-space islands, island overlap, and chaos. 

We aim to address unresolved questions in this area, including how the quasisymmetry helicity and deviations from quasisymmetry impact transport. Recent simulations \citep{2022White,2023White} of resonant AEs in W7-X and the precise QH equilibrium \citep{2022Landreman} have indicated that even a small-amplitude Alfv\'{e}nic perturbation, $\delta B/B \sim 10^{-6}$, can lead to global flattening of the distribution function. This extreme sensitivity to perturbations is postulated to arise because the low magnetic shear of the equilibrium implies low transit frequency shear for passing particles. Here, we study the impact of magnetic shear in more detail by evaluating the island width and drift-harmonic overlap conditions for quasisymmetric configurations. As discussed later, our conclusions do not quantitatively agree with these recent studies, which predict substantive diffusive losses for low AE perturbation amplitudes for equilibria very close to QS. 

The impact of AEs on the fast-ion transport in a stellarator reactor is challenging to compute in practice, requiring knowledge of the saturated mode amplitude. The nonlinear saturation can be obtained from high-fidelity modeling \citep{2017Todo,2014Feher,2017Spong}, which depends on the details of the thermal and fast-ion profiles. Rather than attempt such calculations here, we consider the potential impact of a single Alfv\'{e}nic perturbation on the phase-space integrability and resulting transport in reactor-scale equilibria designed to be close to quasisymmetry. We employ a ``worst-case scenario'' approach, in which an Alfv\'{e}nic perturbation is chosen to strongly resonate with rational periodic passing orbits in the core for several quasisymmetric configurations. We consider several mode amplitudes consistent with experimental measurements and high-fidelity modeling. While physically, such a perturbation should correspond to an AE of the background plasma, we instead let the perturbation be a radially global mode with prescribed mode numbers, frequency, and perturbation amplitude. We assess the impact of the perturbation mode number, amplitude, and residual quasisymmetry error on the resulting transport. Phase-space Poincar\'{e} plots are used to guide the analysis and assess the transition to chaos. While Poincar\'{e} plots have been used to study the structure of phase space for energetic passing particles in the absence of time-dependent perturbations for some stellarators \citep{2022White}, this technique has not yet been applied to assess the impact of AEs on the transport. We evaluate the extent to which this technique provides insight into the transport in configurations with varying deviations from quasisymmetry. 

In Section \ref{sec:gc_equations}, we outline the guiding center equations of motion with Alfv\'{e}nic perturbations. In Section \ref{sec:resonance_theory}, we describe the theory for resonance, island width, and island overlap in QS configurations. Resonance analysis is performed for several equilibria designed to be close to QS in Section \ref{sec:resonance_analysis}, and resonant perturbations are identified. Kinetic Poincar\'{e} plots are employed in Section \ref{sec:kinetic_poincare} to assess the formation of phase-space islands and chaos in the presence of the resonant perturbations. A Monte Carlo guiding center transport analysis is performed in Section \ref{sec:Monte_carlo}. We conclude in Section \ref{sec:conclusions}.

\section{Guiding-center motion in the presence of an Alfv\'{e}nic perturbation}
\label{sec:gc_equations}

The guiding-center motion, $\bm{R}(t)$, is described by the Lagrangian \citep{1983Littlejohn},
\begin{align}
    L(\bm{R},\dot{\bm{R}},v_{\|}) = q\left(\bm{A} + \frac{M v_{\|}}{q B}\bm{B} \right) \cdot \dot{\bm{R}} - \frac{M v_{\|}^2}{2} - \mu B - q\Phi,
    \label{eq:Lagrangian_unperturbed}
\end{align}
where $q$ is the charge, $M$ is the mass, $v_{\|}$ is the velocity in the direction of the magnetic field, and $\mu$ is the magnetic moment. We assume the magnetic field is comprised of an equilibrium field, $\bm{B}_0$, and a shear Alfv\'{e}nic perturbation \citep{1983White},
\begin{align}
    \bm{B} = \bm{B}_0 + \nabla \times \left(\alpha \bm{B}_0 \right),
    \label{eq:perturbed_field_alfvenic}
\end{align}
while the scalar potential vanishes in the equilibrium, $\Phi = \delta \Phi$. With the reduced MHD assumption \citep{1998Kruger}---$k_{\|}/k_{\perp} \ll 1$, where $k_{\|}$ is the characteristic parallel wave number and $k_{\perp}$ is the characteristic perpendicular wave number associated with perturbed quantities---the form of the perturbed field \eqref{eq:perturbed_field_alfvenic} implies that the linear perturbation to the field strength vanishes, $\delta B = 0$. Under the ideal MHD assumption, the corresponding scalar potential must satisfy $\bm{B}_0 \cdot\delta \bm{E} =0$, which implies
\begin{gather}
    \nabla_{\|} \delta \Phi = - B_0\partder{\alpha}{t}.
    \label{eq:mde_alpha_phi}
\end{gather}
The perturbed fields could be computed from a global Alfv\'{e}n eigenmode solver such as AE3D \citep{2010Spong}. For the fundamental studies here, we instead assume a single harmonic perturbation of the form,
\begin{align}
    \delta \Phi = \hat{\Phi}(\psi) \sin(\omega t + m \theta - n \zeta),
    \label{eq:delta_phi}
\end{align}
where $(\psi,\theta,\zeta)$ are Boozer coordinates such that the unperturbed magnetic field is expressed as,
\begin{align}
\left\{
    \begin{array}{l}
      \displaystyle
    \bm{B}_0 = \nabla \psi \times \nabla \theta - \iota(\psi) \nabla \psi \times \nabla \zeta, \\ 
    \bm{B}_0 = G(\psi) \nabla \zeta + I(\psi) \nabla \theta + K(\psi,\theta,\zeta) \nabla \psi,
    \end{array}
    \right .
\end{align}
$m$ is the poloidal mode number, $n$ is the toroidal mode number, and $\omega$ is the frequency. We define the phase variable $\eta = \omega t + m \theta - n \zeta$. 
The corresponding expression for $\alpha$ is then obtained from \eqref{eq:mde_alpha_phi}, which we write schematically as,
\begin{align}
    \alpha = \hat{\alpha}(\psi) \sin(\eta).
    \label{eq:alpha}
\end{align}
We impose $\hat{\Phi}$ and compute $\hat{\alpha}$ so that a magnetic differential equation need not be inverted, which can give rise to singular solutions on rational surfaces. 

Since $M v_{\|}/(qB_0)$ is small in comparison to the characteristic length scale of the equilibrium and $\delta B=0$, the perturbed Lagrangian reads \citep{1985Littlejohn},
\begin{align}
    L(\bm{R},\dot{\bm{R}},v_{\|}) = q\left(\bm{A}_0 + \alpha \bm{B}_0 + \frac{Mv_{\|}}{qB_0} \bm{B}_0 \right) \cdot \dot{\bm{R}} - \frac{M v_{\|}^2}{2} - \mu B_0 - q\delta \Phi. 
\end{align}
The resulting equations of motion are integrated in Boozer coordinates using the SIMSOPT stellarator optimization and modeling package \citep{2021Landreman}.

To compare Alfv\'{e}nic perturbations across configurations with different mode numbers, we will express the strength of the perturbation in terms of its normalized radial magnetic field,
\begin{align}
   \frac{\delta \bm{B} \cdot \nabla \psi}{B_0 |\nabla \psi|} \approx \frac{\nabla \alpha \times \bm{B}_0 \cdot \nabla \psi}{r B_0^2} = \frac{\hat{\alpha} (m G -n I)}{r (G + \iota I)} \cos(\eta).
\end{align}
Here we have assumed that the gradient scale length of the perturbed field is larger than that of the equilibrium. Since $G \gg I $ for stellarator equilibria, we define the parameter $\delta \hat{B}^{\psi} = m \hat{\alpha}/r \sim \delta \bm{B} \cdot \nabla \psi/(B_0 |\nabla \psi|)$ to compare equilibria with respect to the strength of the perturbed radial field. 

\section{Resonance theory}
\label{sec:resonance_theory}

We assume a quasisymmetric equilibrium for which the unperturbed field strength can be expressed as $B_0(\psi,\chi)$ where $\chi = \theta - N \zeta$ is the symmetry angle and $N$ is an integer representing the symmetry helicity. ($N = 0$ for quasiaxisymmetry and $N \ne 0$ for quasihelical symmetry.) Each equilibrium we consider in Section \ref{sec:resonance_analysis} is sufficiently close to quasisymmetry that such an assumption provides valuable insight into the resulting dynamics. 

The radial drift over the unperturbed trajectories is analyzed in the presence of an Alfv\'{e}nic perturbation in Appendix \ref{app:resonance_condition}, generalizing the theory of \cite{1993Mynicka} to quasisymmetric configurations and moderate frequency perturbations. 
At moderate frequencies, the electrostatic potential enters the guiding center equations of motion. Because $\bm{B}_0 \cdot \delta \bm{E} = 0$, particles continue to follow perturbed field lines to lowest order in the gyroradius but experience an additional $\bm{E} \times \bm{B}$ drift. 
Under the assumption of quasisymmetry and stellarator symmetry, the unperturbed equations of motion can be written schematically as,
\begin{align}
    \left\{
    \begin{array}{l}
    \displaystyle
    \dot{\chi} = \omega_{\chi} + \sum_{j\ne 0} \chi_j \cos(j\chi), \\ \displaystyle
    \dot{\zeta} = \omega_{\zeta} + \sum_{j\ne 0} \zeta_j \cos(j\chi),
    \end{array}
    \right. 
    \label{eq:unperturbed_drifts}
\end{align}
where $\omega_{\chi} = \langle \dot{\chi} \rangle$ and $\omega_{\zeta} = \langle \dot{\zeta} \rangle$ are the averaged drifts in the $\chi$ and $\zeta$ directions. The overdot represents a time derivative, and the average is performed over many toroidal transits, $\langle A \rangle = \int_0^T dt \, A /T$. The summations represent the periodic contributions from the drifts. 

For an Alfv\'{e}nic perturbation of the form \eqref{eq:delta_phi} to provide a net radial drift, a resonance condition must be satisfied,
\begin{align}
    \Omega_l = (m+l)\omega_{\chi} - (n - Nm)\omega_{\zeta} + \omega = 0.
    \label{eq:resonance_condition}
\end{align}
As discussed in Appendix \ref{app:resonance_condition}, 
the integer $l$ arises due to coupling through the $\chi$-dependence of the magnetic drifts. 
If the drift dynamics \eqref{eq:unperturbed_drifts} is dominated by a particular cosine harmonic with integer $j'$, then $l$ is assumed to be an integer multiple of $j'$. To simplify the analysis, the $j' = 1$ harmonic of the field strength is assumed to be dominant, which holds near the magnetic axis \citep{1991Garren} and for the equilibria of interest. More general expressions are provided in Appendix \ref{app:resonance_condition}. Under the assumption that the characteristic frequencies are approximately flux functions, the full island width associated with a given resonance is given by,
\begin{align}
    w^{\psi}_{l} = 2\sqrt{\left | \frac{\psi_{l}}{\Omega_{l}'(\psi)} \right |} \approx 2\sqrt{\left | \frac{\psi_{l\ne 0}}{(m + l) \omega_{\theta}'(\psi)} \right |},
    \label{eq:island_width_basic}
\end{align}
where $\psi_l$ is a cosine harmonic of the radial perturbed drift and we have made the approximation $h'(\psi)\approx \omega_{\theta}'(\psi)/\omega_{\zeta}$. The perturbed radial drift, $\delta \dot{\psi}$, can be evaluated along the unperturbed trajectory and expressed as a cosine series in $\cos(\Omega_l t + \eta^0)$ and $\cos(\Omega_l t + \eta^0 \mp \chi^0)$ with coefficients given by $\psi_l^0$ and $\psi_l^{\pm}$; see \eqref{eq:net_radial_drift}. Here $\eta^0 = m \theta^0 - n \zeta^0$ is the initial phase. The scaling of these coefficients with the magnitude of the magnetic drifts, mode numbers, and perturbed radial field is summarized as,
\begin{align}
\left \{ \renewcommand{\arraystretch}{1.7}{\begin{array}{l} \displaystyle
\psi_{0}^0 \sim (\iota - \omega_{\theta}/\omega_{\zeta}) J_0(\eta_{1}) \delta \hat{B}^{\psi}, \\  \displaystyle
\psi_l^0 \sim J_l(\eta_{1})\delta \hat{B}^{\psi}, \\  \displaystyle
\psi_{l}^{\pm} \sim J_{l\pm 1}(\eta_{1}) \zeta_{1} \delta \hat{B}^{\psi}, 
\end{array}} \right. 
\label{eq:psi_cs}
\end{align}
where $J_l$ are the Bessel functions of the first kind, $\eta_{1} = \left(m\chi_{1}- (n-Nm)\zeta_{1}\right)/\omega_{\chi}$ with $\chi_1$ and $\zeta_1$ defined through \eqref{eq:unperturbed_drifts}, and $\omega_{\theta} = \langle \dot{\theta} \rangle$. The full expressions for general $j'$ are provided in Appendix \ref{app:resonance_condition}.

When considering the dependence of the island width on the magnetic drifts in the small argument limit of the Bessel functions, the most significant radial transport will arise from $\psi_0^0$, $\psi_{\pm 1}^0$, or $\psi_{\mp 1}^{\pm }$. In the limit of large mode numbers, $\psi_{\pm 1}^0$ will dominate due to its dependence on $\eta_1$. 
Considering the small argument limit of the Bessel function, the island width scales with $\sqrt{\eta_1^{|l|}/(m+1)}$ for fixed $\delta \hat{B}^{\psi}$, thus increasing the island width for quasihelical configurations due to the dependence of $\eta_1$ on the helicity $N$. 
Given the scaling of $\eta_1$ with the mode numbers, the island width will roughly scale independently of $m$ for $|l| =1$ but will scale with $m^{(|l|-1)/2}$ for $|l|>1$.
Finally, the island width decreases strongly with increasing $|l|$. 

Defining the passing orbital helicity as $h = \omega_{\theta}/\omega_{\zeta}$, resonances occur where:
\begin{align}
     h = \frac{n - Nm - \omega/\omega_{\zeta}}{m + l}+N.
     \label{eq:orbit_helicity}
\end{align}
For a given primary resonance $l$ at $h = h_0$, additional sideband resonances may be excited for other drift harmonics $l'$ corresponding to neighboring periodic orbits. The spacing between the $l$ and $l+1$ resonances is given by,
\begin{align}
    \left(\Delta \psi\right)_l = \frac{1}{h'(\psi)}\frac{h_0 - N}{m + l+1}.
    \label{eq:resonance_spacing}
\end{align}
The ratio between the island width and the resonance spacing
\begin{align}
    \frac{w_l^{\psi}}{\left(\Delta \psi\right)_l} \approx \frac{m + l + 1}{(h_0 - N)\omega_{\zeta}}\sqrt{\left | \frac{\psi_l \omega_{\theta}'(\psi)}{m + l} \right |}
\end{align}
provides a conservative estimate for the island overlap criterion, $w_l^{\psi}/\left(\Delta \psi\right)_l \gtrsim 1$.
Given the scaling of $\psi_l$ \eqref{eq:psi_cs}, the potential for island overlap increases with $m$ and decreases with $N$ for fixed $\delta \hat{B}^{\psi}$. In this way, quasihelical configurations are advantageous for preventing the transition to phase-space chaos. Finally, while shear in the transit frequency, $\omega_{\theta}'(\psi)$, reduces individual island widths, it promotes island overlap if multiple resonances are present. 

\section{Resonance analysis}
\label{sec:resonance_analysis}

\begin{figure}
    \centering
    \begin{subfigure}{0.49\textwidth}
    \includegraphics[width=1.0\textwidth]{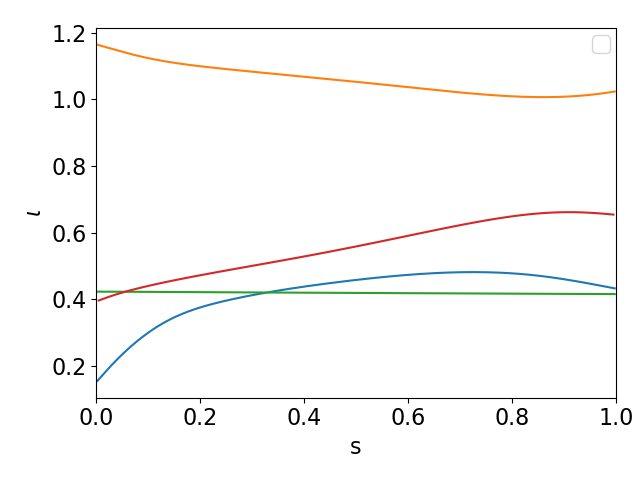}
    \caption{}
    \end{subfigure}
    \begin{subfigure}{0.49\textwidth}
    \includegraphics[width=1.0\textwidth]{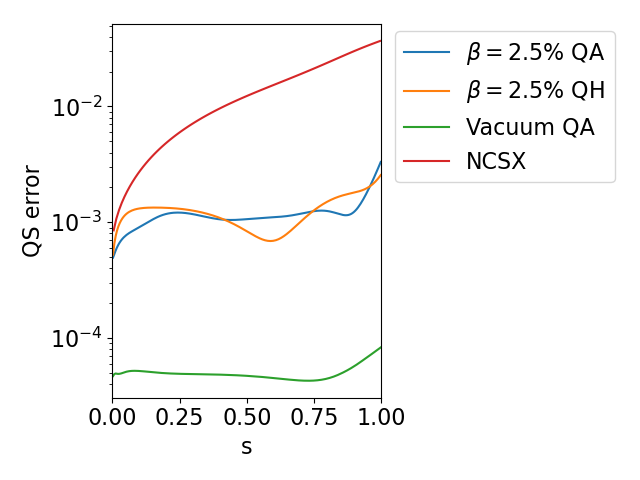}
    \caption{}
    \end{subfigure}
    \caption{(a) Rotational transform and (b) quasisymmetry error \eqref{eq:qs_error} profiles for the four equilibria under consideration.}
    \label{fig:qs_iota}
\end{figure}

We consider four equilibria optimized to be close to quasisymmetry: $\beta = 2.5\%$ quasihelical (QH) and quasiaxisymmetric (QA) equilibria with self-consistent bootstrap current \citep{2022Landremanb}, a vacuum equilibrium with precise levels of quasiaxisymmetry \citep{2022Landreman}, and the quasiaxisymmetric NCSX li383 equilibrium \citep{2003Koniges,2002Mynick}. Each equilibrium considered is scaled to the minor radius (1.70 m) and the field strength (5.86 T) of ARIES-CS \citep{2008Najmabadi}. The rotational transform profiles and quasisymmetry error, \begin{equation}
    f_{QS}(s) = \frac{\sqrt{\sum_{Mn\ne Nm} \left(B_{m,n}^c(s)\right)^2}}{\sqrt{\left(B_{0,0}^c(s)\right)^2}},
    \label{eq:qs_error}
\end{equation}
are shown in Figure \ref{fig:qs_iota},
where the unperturbed field strength in Boozer coordinates is $B_0(s,\theta,\zeta) = \sum_{m,n}B_{m,n}^c(s) \cos(m \theta - n \zeta)$. 

We first identify the low-order periodic passing orbits present in the equilibrium to determine the impact of a potential resonant perturbation on the guiding center losses. Fusion-born alpha particles are initialized with equally-spaced pitch angle and radius values and followed for 500 toroidal transits. The net change in the poloidal angle, $\Delta \theta$, and transit time, $\Delta t$, are computed for each toroidal transit. In the integrable case, the average of the characteristic frequencies over many transits, denoted $\omega_{\zeta} = 2\pi/\langle \Delta t \rangle$, and $\omega_{\theta} = \langle \Delta \theta \rangle/\langle \Delta t \rangle$, converges quickly with respect to the number of toroidal transits \citep{2016Das}. 
In this way, we obtain $\omega_{\zeta}$ and $\omega_{\theta}$ in the two-dimensional space $(s,\mu)$, where $s = \psi/\psi_0$ is the normalized toroidal flux and $\psi_0$ is the value of the toroidal flux on the boundary. All nonsymmetric modes are artificially suppressed for this frequency analysis so that all trajectories lie on KAM surfaces. 

For the subsequent analysis, resonant perturbations are intentionally imposed, satisfying the condition \eqref{eq:resonance_condition} for a resonant surface near mid-radius. As many potential Alfv\'{e}nic perturbations exist that resonate with a given periodic orbit, a comparison is made between $m = 1$, 15, and 30 perturbations. The perturbation with the highest mode number is chosen such that $m \approx a/\rho_{EP}$ for the ARIES-CS reactor parameters, where $a$ is the effective minor radius and $\rho_{EP}$ is the energetic particle orbit width. 
With this choice, the radial width of the AE eigenstructure is predicted to be comparable to the EP orbit width, and the growth rate is maximized \citep{2014Gorelenkov}. 
The smaller values of $m$ are more typical for current experiments such as LHD \citep{2017Varela}. We choose perturbation amplitudes in the range $\delta \hat{B}^{\psi} \sim 10^{-4}-10^{-3}$, consistent with LHD modeling \citep{2013Nishimura} and experimental measurements from TFTR and \citep{1997Nazikian} and NSTX \citep{2013Crocker}. In analysis of the impact of TAEs on guiding center confinement in tokamaks, alpha orbit stochasticity is typically present for $\delta \hat{B}^{\psi} \sim 10^{-3}$, while for $\delta \hat{B}^{\psi} < 10^{-4}$ the losses are insignificant \citep{1992Sigmar}. Although the eigenfunction typically depends on the mode number, we choose a radially uniform mode structure for a conservative analysis.

Given a $\omega_{\theta}/\omega_{\zeta} = p/q$ periodic orbit and poloidal mode number $m$, the resonant wave frequency satisfying \eqref{eq:resonance_condition} can be expressed as $\omega/\omega_{\zeta} = p'/q$ for some integer $p'$. The toroidal mode number $n$ and the frequency parameter $p'$ are chosen to satisfy the $l = 1$ resonance condition \eqref{eq:resonance_condition}, given that the $l = \pm 1$ resonance is predicted to give rise to the most substantial radial transport as described in Section \ref{sec:resonance_theory}. The value of $n$ is chosen to yield small magnitudes of $|p'|$, corresponding with lower-frequency perturbations. The mode parameters are summarized in Table \ref{tab:mode_parameters}.

\begin{table}
    \centering
    \begin{subtable}{0.49\textwidth}
    \centering 
    \begin{tabular}{|c|c|c|c|c|c|c|} 
        $m$ & $n$ & $\omega/\omega_{\zeta}$  & $\omega_{\theta}/\omega_{\zeta}$ & $\omega$ [kHz] \\ \hline 
        1 & 1 & 1/9 & 4/9 & 136 \\ 
        15 & 7 & -1/9 & 4/9 & -136 \\ 
        30 & 14 & 2/9 & 4/9 & 272\\ 
    \end{tabular}
    \caption{$\beta = 2.5\%$ QA ($N = 0$)}
    \end{subtable}
    \begin{subtable}{0.49\textwidth}
    \centering 
    \begin{tabular}{|c|c|c|c|c|c|c|} 
        $m$ & $n$ & $\omega/\omega_{\zeta}$ & $\omega_{\theta}/\omega_{\zeta}$ & $\omega$ [kHz]\\ \hline 
        1 & -2 &  2/14 & -15/14 & 133 \\ 
        15 & 13 & 2/14 & -15/14 & 133 \\ 
        30 & 29 &  3/14 & -15/14 & 200
    \end{tabular}
    \caption{$\beta = 2.5\%$ QH ($N = 4$)}
    \end{subtable}
    \begin{subtable}{0.49\textwidth}
    \centering 
    \begin{tabular}{|c|c|c|c|c|c|c|} 
        $m$ & $n$ &  $\omega/\omega_{\zeta}$ & $\omega_{\theta}/\omega_{\zeta}$ & $\omega$ [kHz] \\ \hline 
        1 & 1 &  2/12 & 5/12 & 199 \\ 
        15 & 7 &  4/12 & 5/12 & 399 \\ 
        30 & 13 &  1/12 & 5/12 & 997 
    \end{tabular}
    \caption{Vacuum QA ($N = 0$)}
    \end{subtable}
    \begin{subtable}{0.49\textwidth}
    \centering 
    \begin{tabular}{|c|c|c|c|c|c|} 
        $m$ & $n$ &  $\omega/\omega_{\zeta}$ & $\omega_{\theta}/\omega_{\zeta}$ & $\omega$ [kHz] \\ \hline 
        1 & 1 & -1/5 & 3/5 & -338 \\ 
        15 & 10 & 2/5 & 3/5 & 676 \\ 
        30 & 19 & 2/5 & 3/5 & 676
    \end{tabular}
    \caption{NCSX ($N = 0$)}
    \end{subtable}
    \caption{Alfv\'{e}nic perturbation mode parameters chosen to satisfy the resonance condition \eqref{eq:resonance_condition}. 
    }
    \label{tab:mode_parameters}
\end{table}

\begin{figure}
    \centering
    \begin{subfigure}{0.49\textwidth}
    \includegraphics[width=1.0\textwidth]{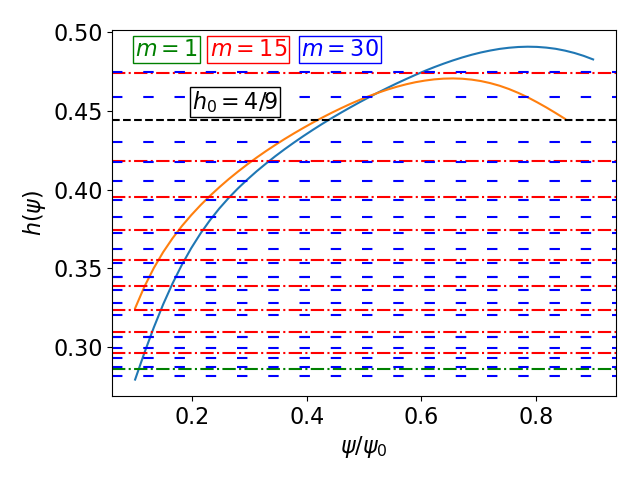}
    \caption{$\beta = 2.5\%$ QA}
    \end{subfigure}
    \begin{subfigure}{0.49\textwidth}
    \includegraphics[width=1.0\textwidth]
    {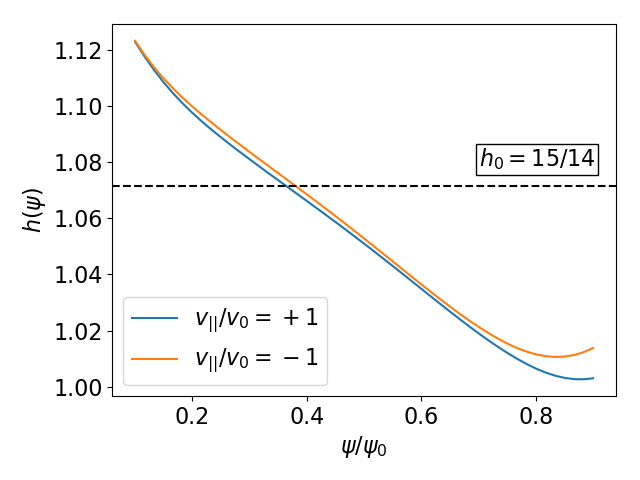}
    \caption{$\beta = 2.5\%$ QH}
    \end{subfigure} 
    \begin{subfigure}{0.49\textwidth}
    \includegraphics[width=1.0\textwidth]{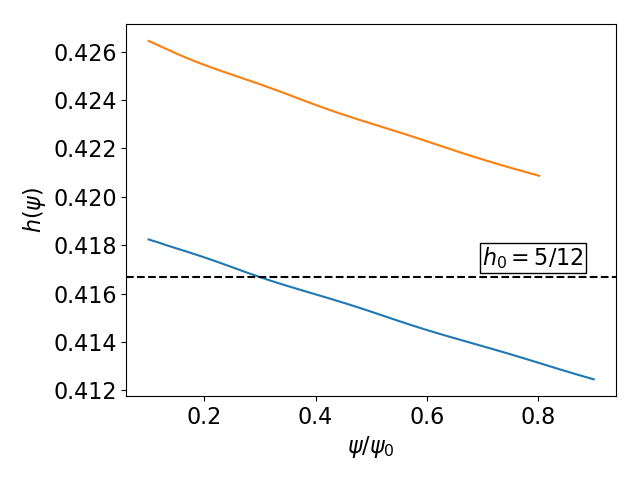}
    \caption{Vacuum QA}
    \end{subfigure}
    \begin{subfigure}{0.49\textwidth}
    \includegraphics[width=1.0\textwidth]{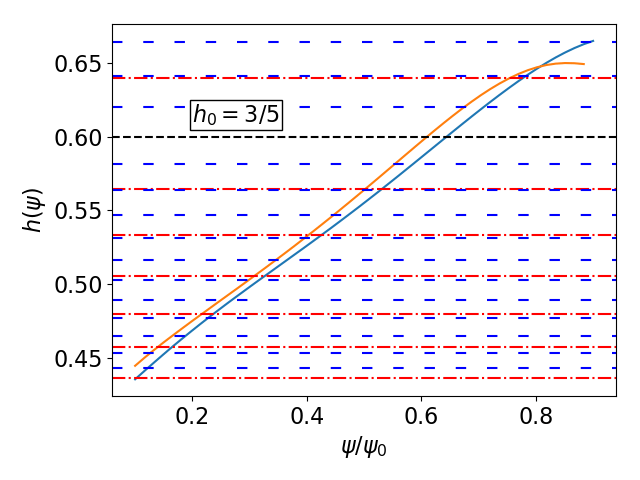}
    \caption{NCSX}
    \end{subfigure}
    \caption{The characteristic orbit helicity, $h = \omega_{\theta}/\omega_{\zeta}$, is computed for co-passing ($v_{\|}/v_0 = +1$, blue) and counter-passing orbits ($v_{\|}/v_0 = -1$, orange). A low-order periodic orbit is selected near the mid-radius, denoted by the horizontal dashed black line. Alfv\'{e}nic perturbations with several mode numbers $m$ are chosen to resonate with this orbit periodicity. Sideband resonances are excited in the $\beta = 2.5\%$ QA and NCSX equilibria due to the angular dependence of the drifts, denoted by the colored horizontal lines. Because of the increased distance between resonances, no sidebands are excited for in the vacuum QA or $\beta = 2.5\%$ QH equilibria for the mode numbers chosen.}
    \label{fig:h_0_QA_QH}
\end{figure}

For each configuration, the profile of the effective orbit helicity $h$ \eqref{eq:orbit_helicity} is shown in Figure \ref{fig:h_0_QA_QH} for co- and counter-passing orbits (blue and orange, respectively). 
While the helicity profile does not change substantially between co- and counter-passing orbits, the resonance condition \eqref{eq:orbit_helicity} is modified since the sign and magnitude of $\omega_{\zeta}$ differs. Although the choice of pitch angle will impact the resonance condition, for the following calculations, we focus on co-passing orbits, $v_{\|}/v_0 = +1$. 
The horizontal black lines in the figures indicate the chosen resonant surface for the co-passing orbits. For the $\beta = 2.5\%$ QA and NCSX equilibria, the 
nearby resonant surfaces corresponding to other values of $l$ are indicated by the colored horizontal lines. As $m$ increases, the resonance spacing decreases according to \eqref{eq:resonance_spacing}. For the vacuum QA equilibrium, the magnetic shear is very low, see Figure \ref{fig:qs_iota}, and no sideband resonances exist. Although the shear is moderate for the $\beta = 2.5\%$ QH equilibrium, sideband resonances are not present due to the factor of $N$ in the resonance spacing expression.

\section{Kinetic Poincar\'{e} plots}
\label{sec:kinetic_poincare}

Given a magnetic field with exact quasisymmetry, a kinetic Poincar\'{e} plot can be constructed analogously to the case of axisymmetric magnetic fields \citep{1992Sigmar}.
Given a quasisymmetric field $B_0(\psi,\chi=\theta-N\zeta)$ in the absence of a perturbation, the canonical angular momentum,
\begin{align}
P_{\zeta} = \left(G + N I\right) \left(\frac{m v_{\|}}{B_0} + q \alpha \right) + q \left( N\psi  - \psi_P \right),
\end{align}
and energy,
\begin{align}
E = \frac{m v_{\|}^2}{2} + \mu B_0 + q \delta \Phi,
\end{align}
are conserved. When an Alfv\'{e}nic perturbation is applied with single mode numbers $n$ and $m$, neither $P_{\zeta}$ nor $E$ are conserved, but a conserved quantity is obtained by moving with the wave frame,
\begin{align}
\overline{E}_n = \left(n - N m\right)E - \omega P_{\zeta}.
\label{eq:E_n}
\end{align}
    Given the velocity-space parameters---$\mu$, $\overline{E}_n$, and $\text{sign}(v_{\|})$---and specification of the position in space and time---$t$, $s$ , $\theta$, and $\zeta$---equation \eqref{eq:E_n} provides a nonlinear equation for $v_{\|}$ \citep{1992Hsu}. 

Given the form for the phase factor of the perturbation \eqref{eq:delta_phi}, the resulting equations of motion depend on $(s,\chi,\zeta - \overline{\omega}_n t, v_{\|})$, where $\overline{\omega}_n = \omega/(n - N m)$. The resulting motion is generally 4D. If purely passing particles are considered, then $\text{sign}\left(v_{\|}\right)$ is fixed, $v_{\|}$ can be computed from the other coordinates as described above, and the resulting motion becomes 3D.
For a fixed value of $\overline{E}_n$ and $\mu$, a kinetic Poincar\'{e} map $M(s,\chi) \rightarrow (s',\chi')$ is constructed by moving with the wave frame to eliminate the time dependence. Guiding center trajectories are followed until they intersect a plane of constant $\zeta - \overline{\omega}_nt$. 
Therefore, the resulting Poincar\'{e} section is 2D. 

When quasisymmetry is broken in the equilibrium, $\overline{E}_n$ is no longer precisely conserved. Nonetheless, if the quasisymmetry errors are sufficiently small, the kinetic Poincar\'{e} analysis provides insight into the resulting transport. The non-quasisymmetric modes are artificially suppressed when constructing the kinetic Poincar\'{e} plots to aid the analysis. The impact of quasisymmetry-breaking modes will be assessed in the following Section.

\begin{figure}
    \centering
    \begin{subfigure}{0.32\textwidth}
    \includegraphics[trim=0.5cm 0.2cm 1.3cm 1.2cm,clip,width=1.0\textwidth]{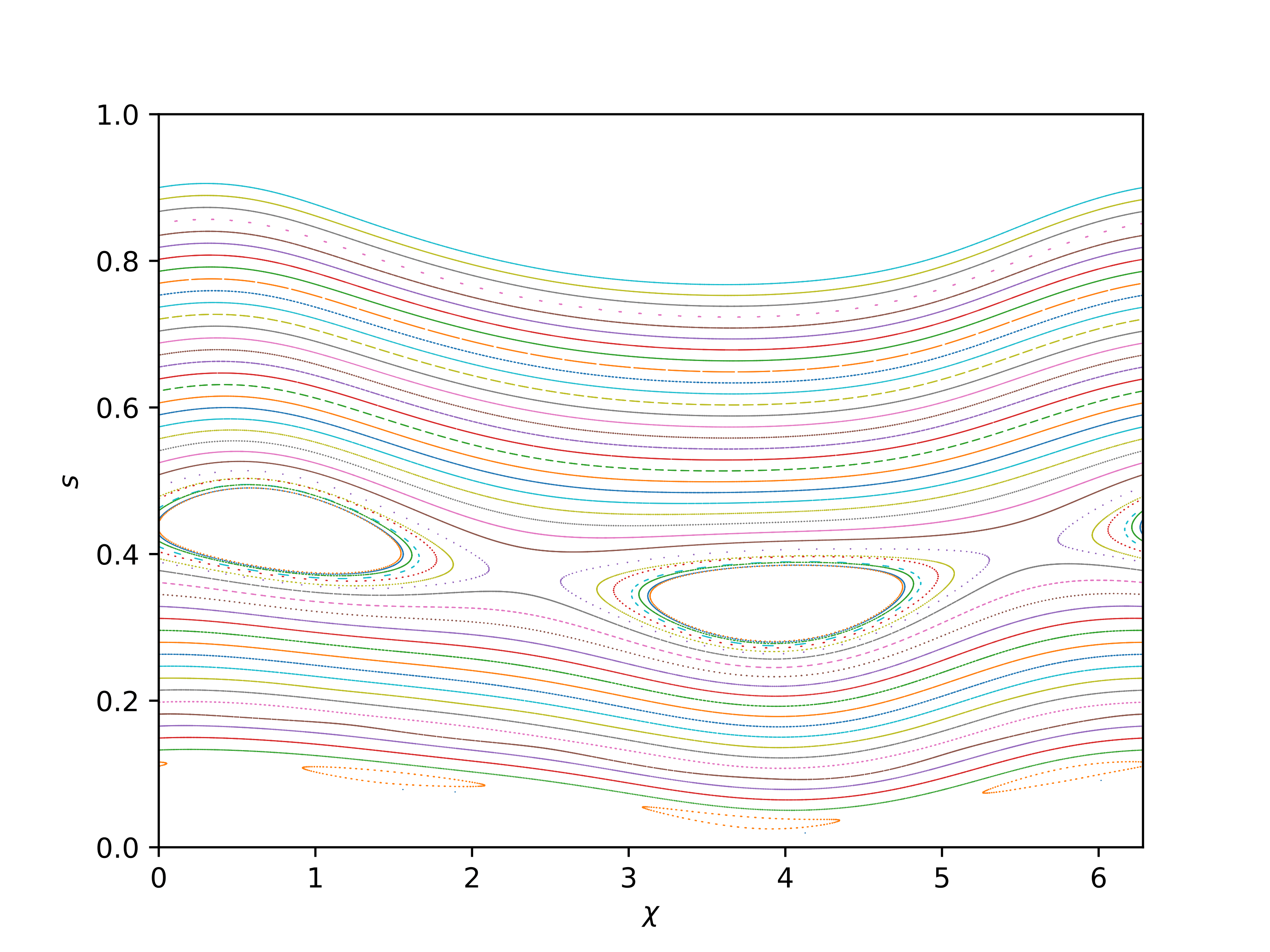}
    \caption{$m = 1$, $\beta = 2.5\%$ QA}
    \end{subfigure}
    \begin{subfigure}{0.32\textwidth}
    \includegraphics[trim=0.5cm 0.2cm 1.3cm 1.2cm,clip,width=1.0\textwidth]{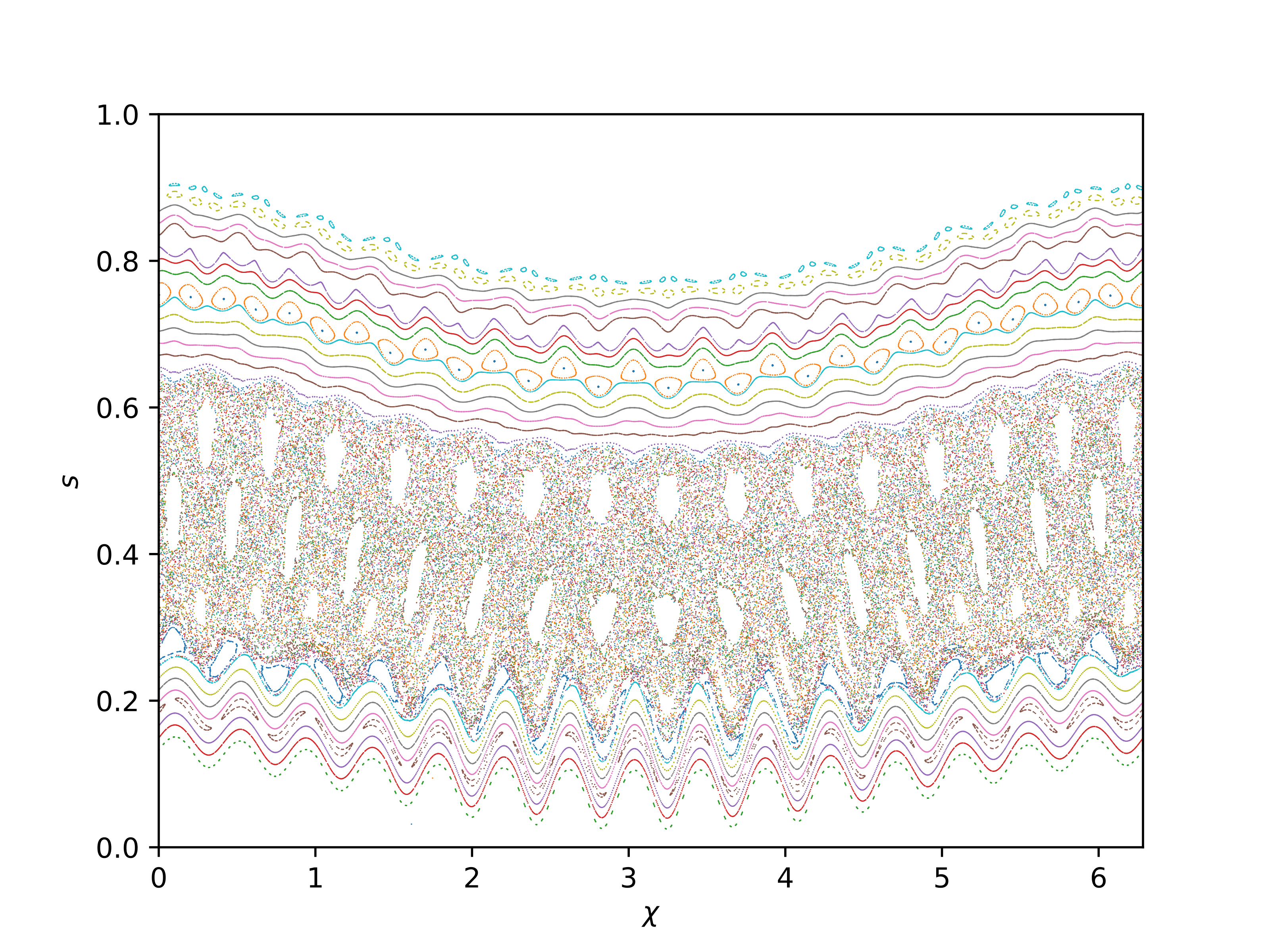}
    \caption{$m = 15$, $\beta = 2.5\%$ QA}
    \end{subfigure}
    \begin{subfigure}{0.32\textwidth}
    \includegraphics[trim=0.5cm 0.2cm 1.3cm 1.2cm,clip,width=1.0\textwidth]{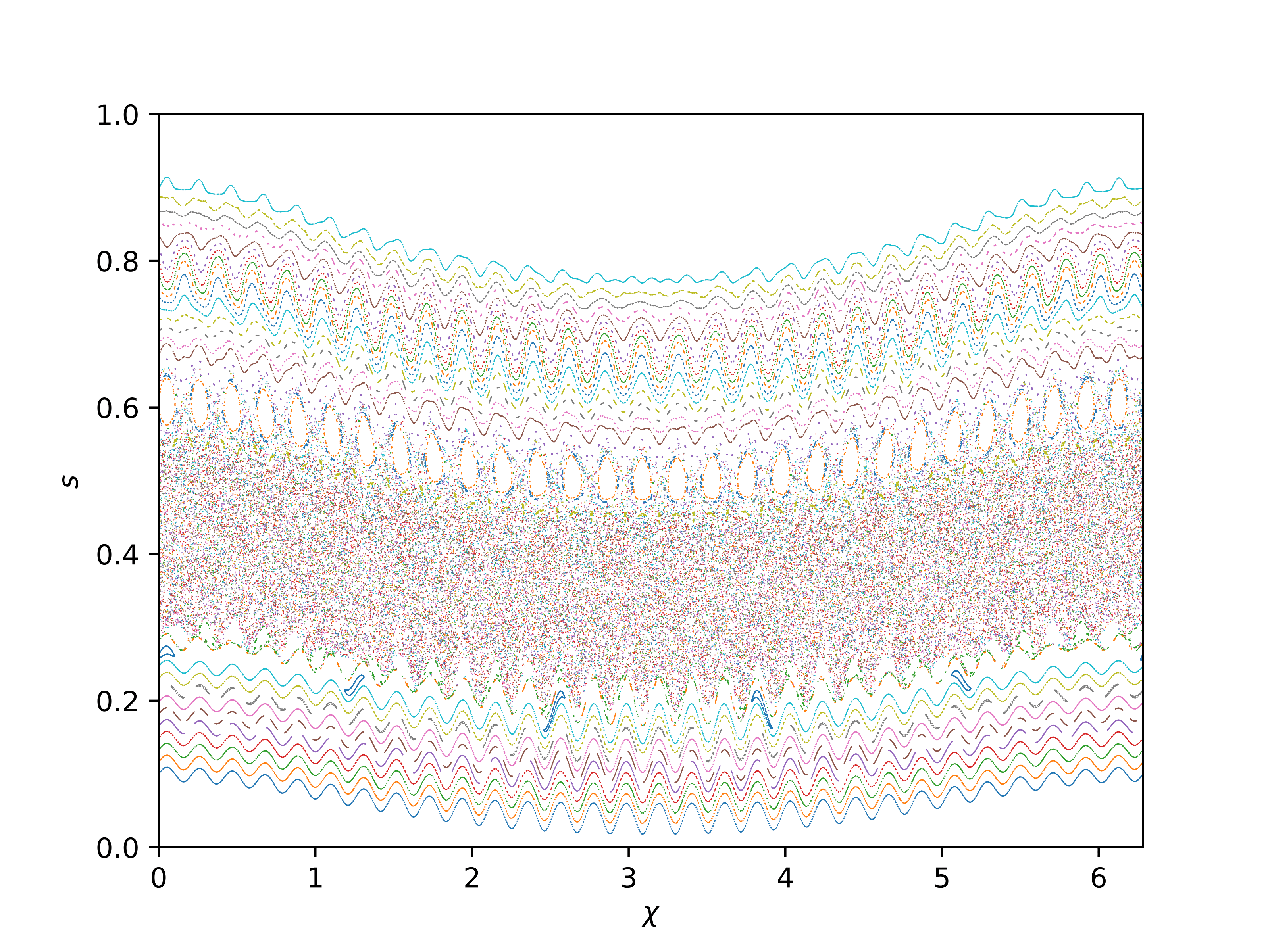}
    \caption{$m = 30$, $\beta = 2.5\%$ QA}
    \end{subfigure}
    \begin{subfigure}{0.32\textwidth}
    \includegraphics[trim=0.5cm 0.2cm 1.3cm 1.2cm,clip,width=1.0\textwidth]{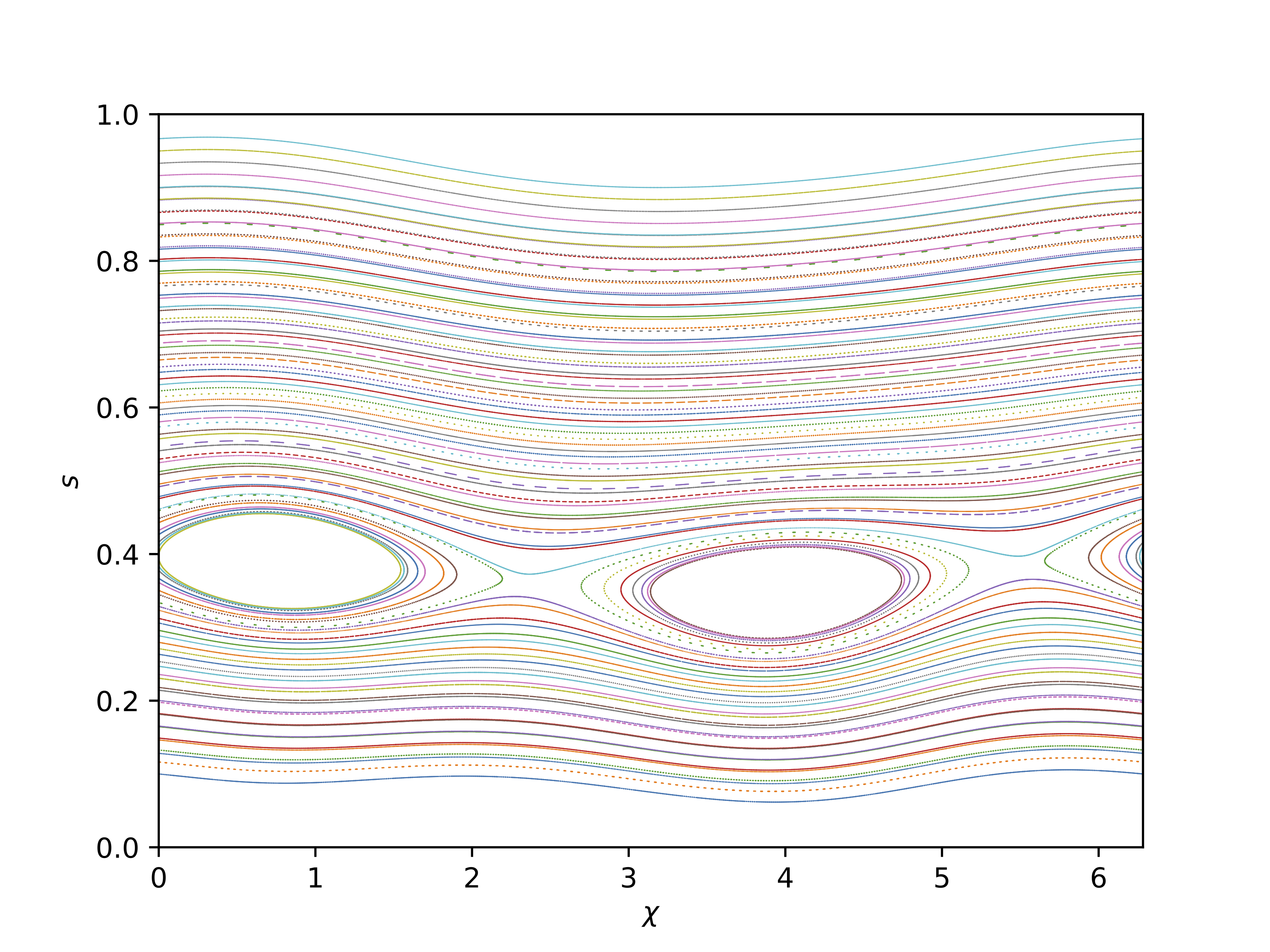}
    \caption{$m = 1$, $\beta = 2.5\%$ QH}
    \end{subfigure}
    \begin{subfigure}{0.32\textwidth}
    \includegraphics[trim=0.5cm 0.2cm 1.3cm 1.2cm,clip,width=1.0\textwidth]{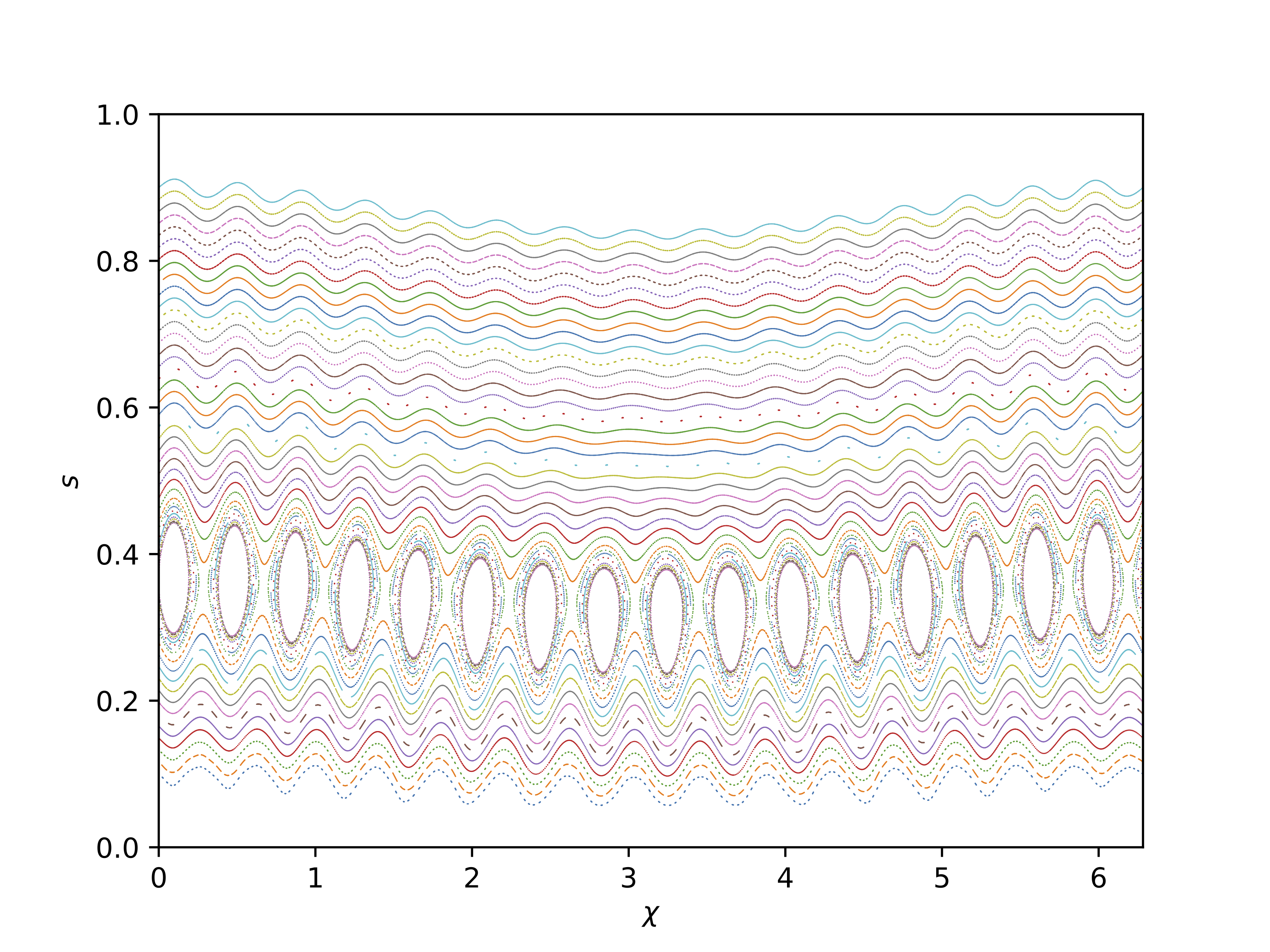}
    \caption{$m = 15$, $\beta = 2.5\%$ QH}
    \end{subfigure}
    \begin{subfigure}{0.32\textwidth}
    \includegraphics[trim=0.5cm 0.2cm 1.3cm 1.2cm,clip,width=1.0\textwidth]{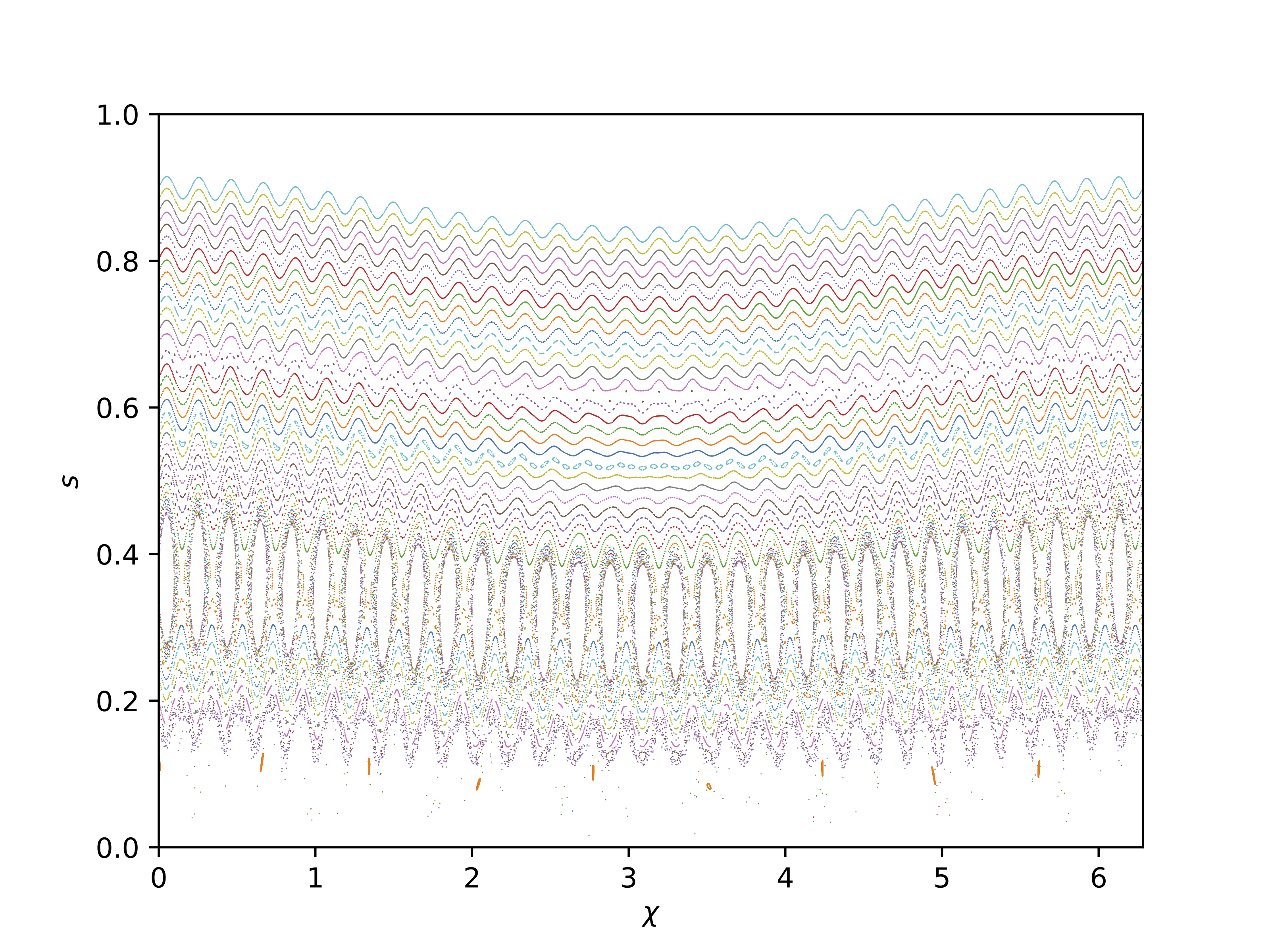}
    \caption{$m = 30$, $\beta = 2.5\%$ QH}
    \end{subfigure}
    \begin{subfigure}{0.32\textwidth}
    \includegraphics[trim=0.5cm 0.2cm 1.3cm 1.2cm,clip,width=1.0\textwidth]{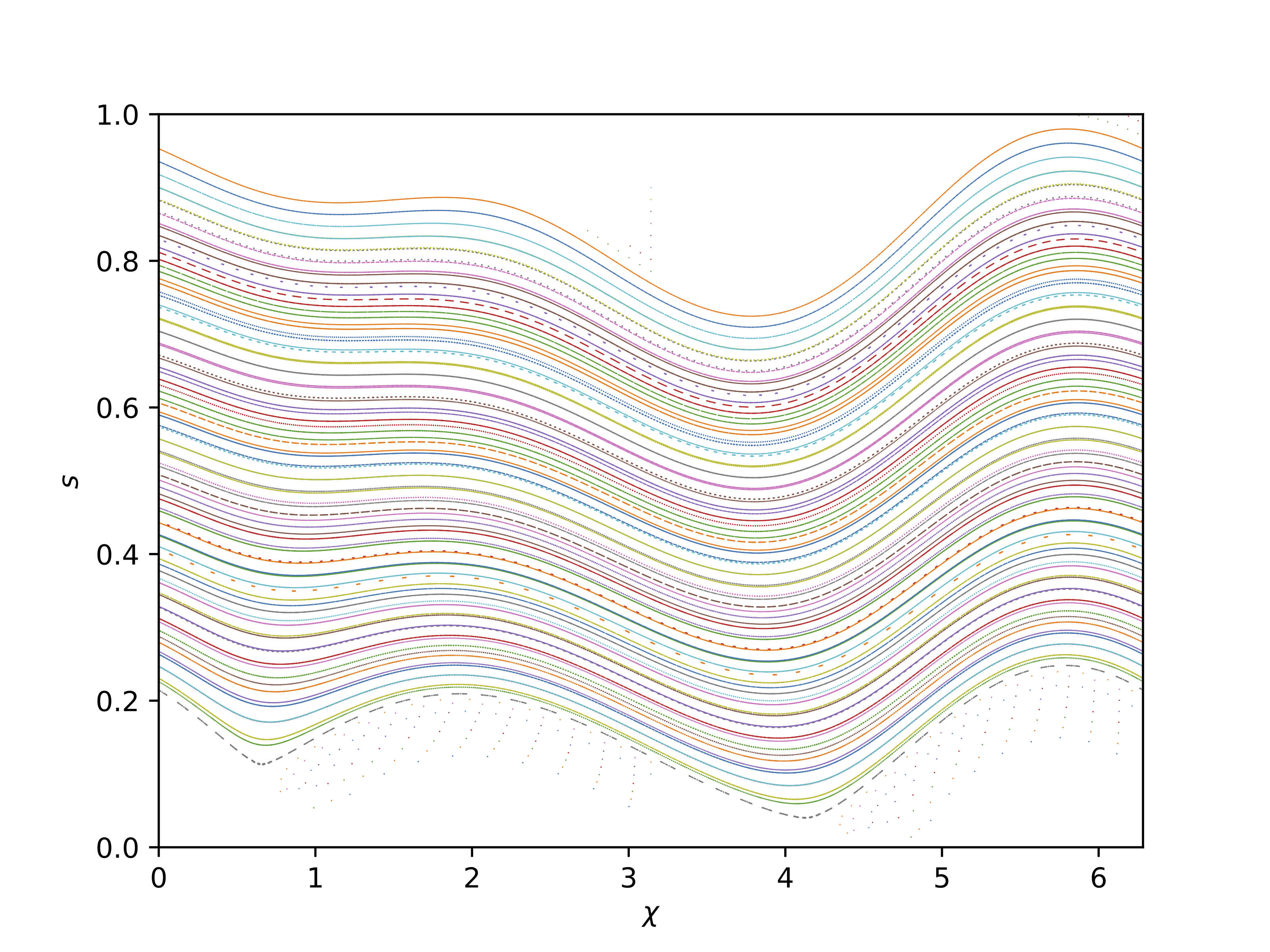}
    \caption{$m = 1$, vacuum QA}
    \label{eq:m_1_vacuum_poinc}
    \end{subfigure}
    \begin{subfigure}{0.32\textwidth}
    \includegraphics[trim=0.5cm 0.2cm 1.3cm 1.2cm,clip,width=1.0\textwidth]{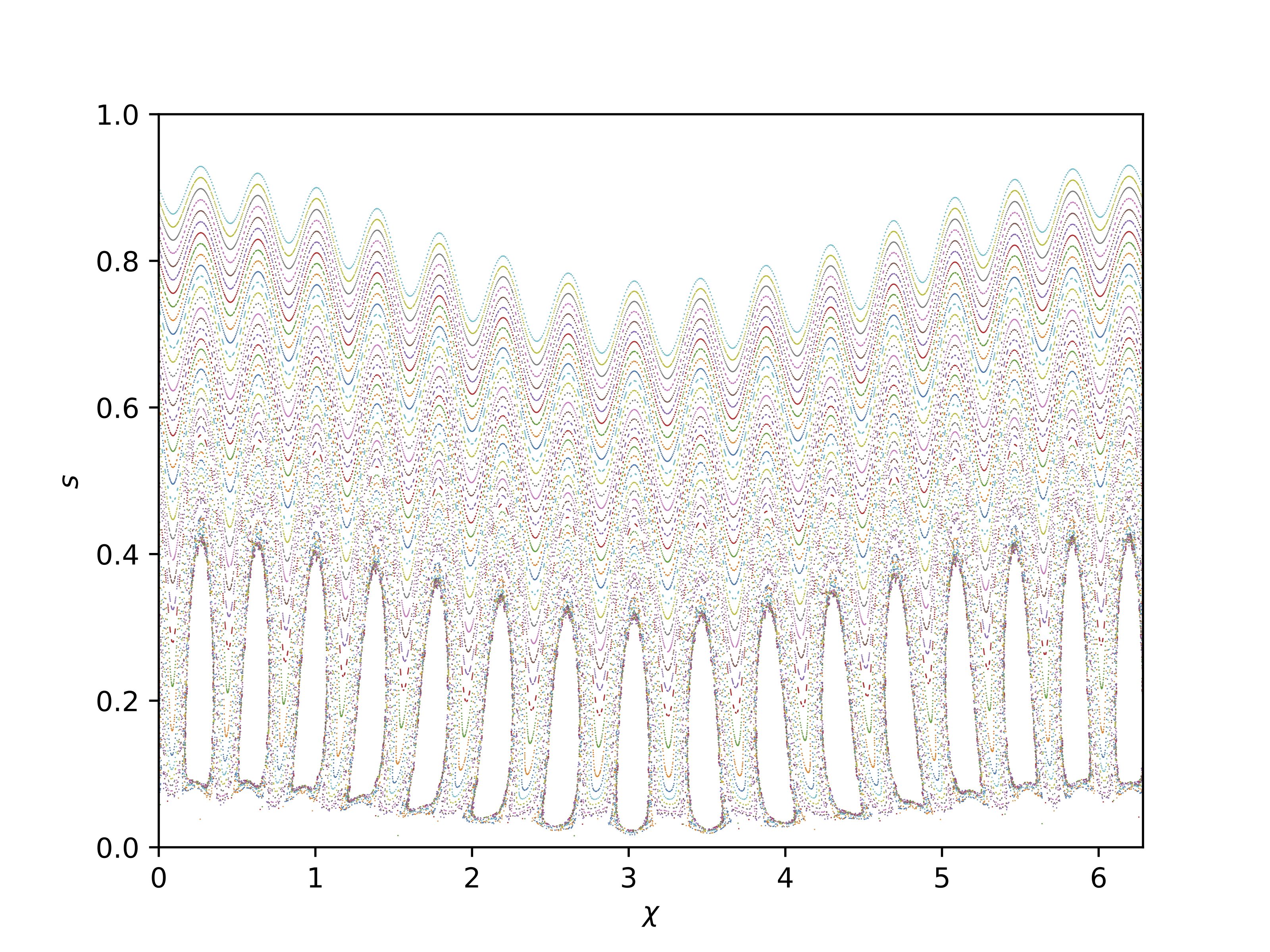}
    \caption{$m = 15$, vacuum QA}
    \end{subfigure}
    \begin{subfigure}{0.32\textwidth}
    \includegraphics[trim=0.5cm 0.2cm 1.3cm 1.2cm,clip,width=1.0\textwidth]{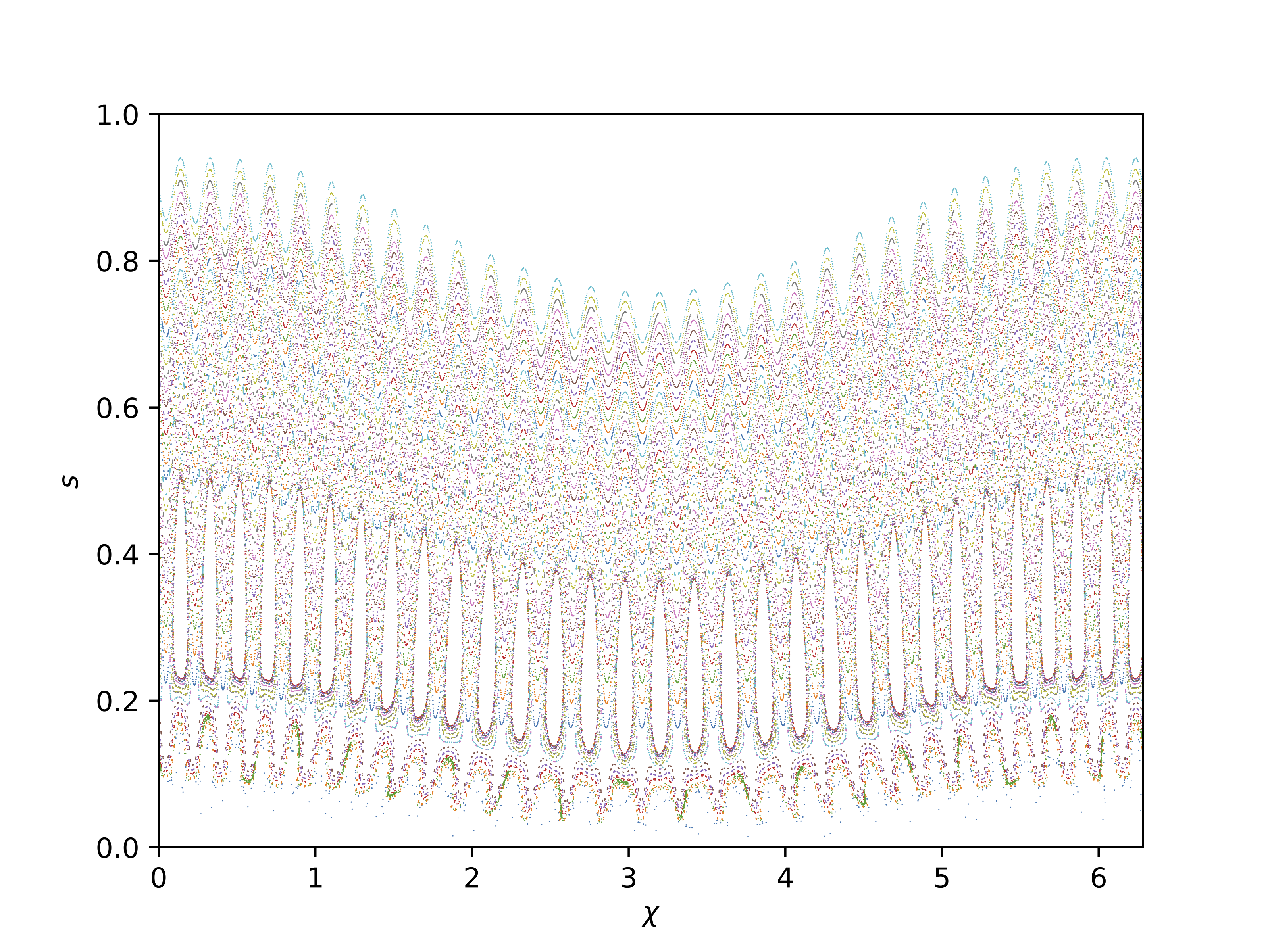}
    \caption{$m = 30$, vacuum QA}
    \end{subfigure}
    \begin{subfigure}{0.32\textwidth}
    \includegraphics[trim=0.5cm 0.2cm 1.3cm 1.2cm,clip,width=1.0\textwidth]{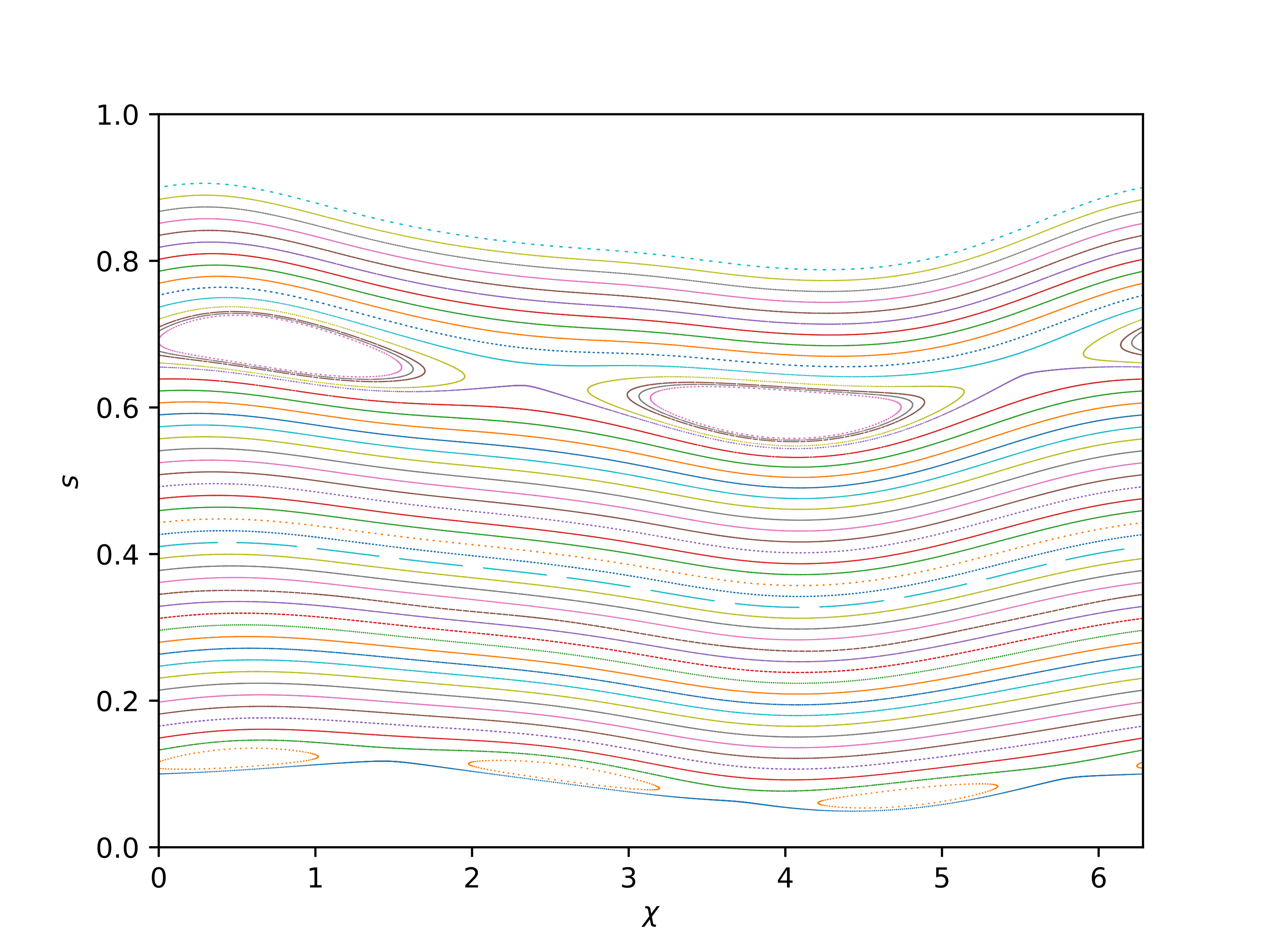}
    \caption{$m = 1$, NCSX}
    \end{subfigure}
    \begin{subfigure}{0.32\textwidth}
    \includegraphics[trim=0.5cm 0.2cm 1.3cm 1.2cm,clip,width=1.0\textwidth]{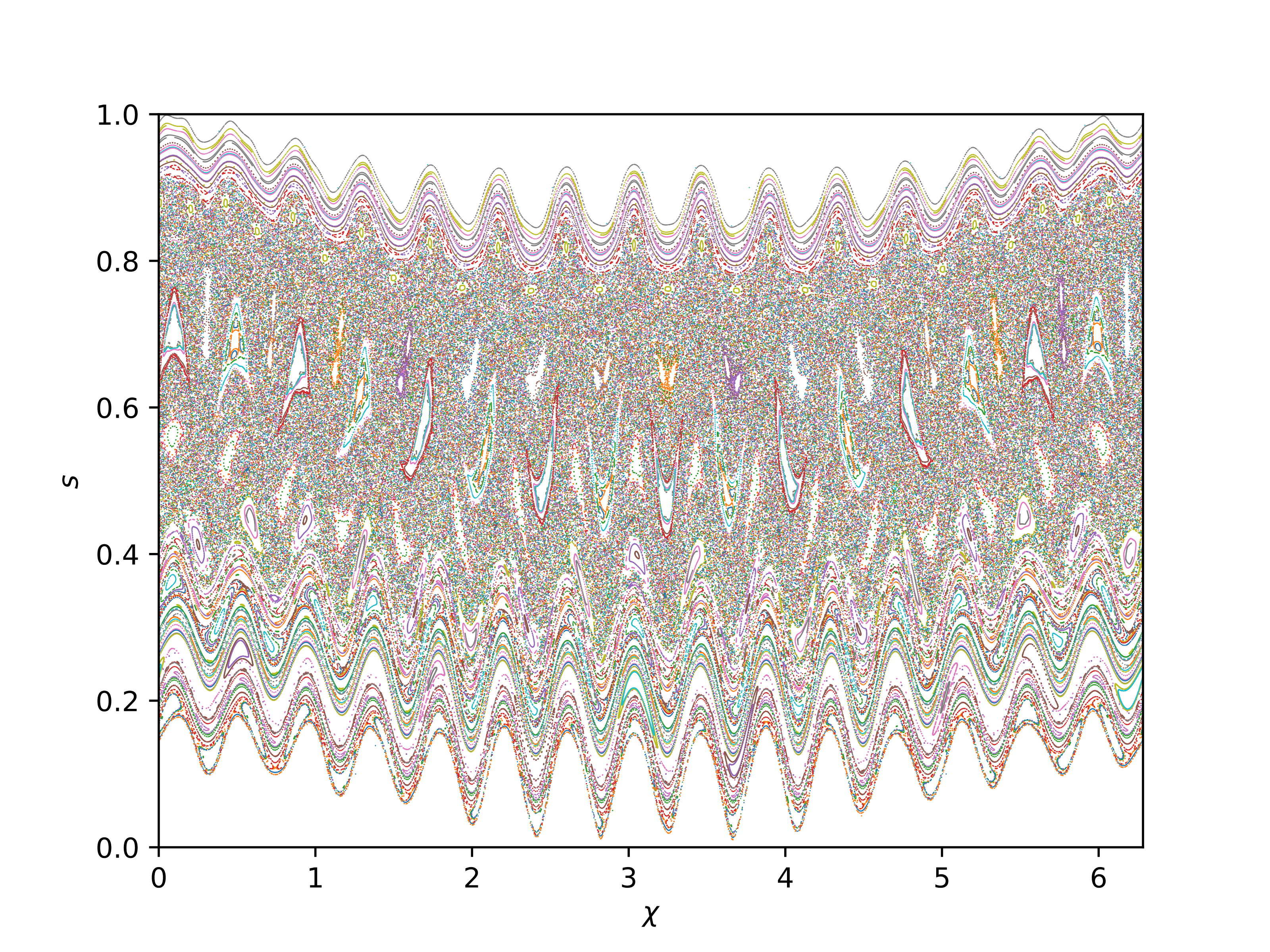}
    \caption{$m = 15$, NCSX}
    \end{subfigure}
    \begin{subfigure}{0.32\textwidth}
    \includegraphics[trim=0.5cm 0.2cm 1.3cm 1.2cm,clip,width=1.0\textwidth]{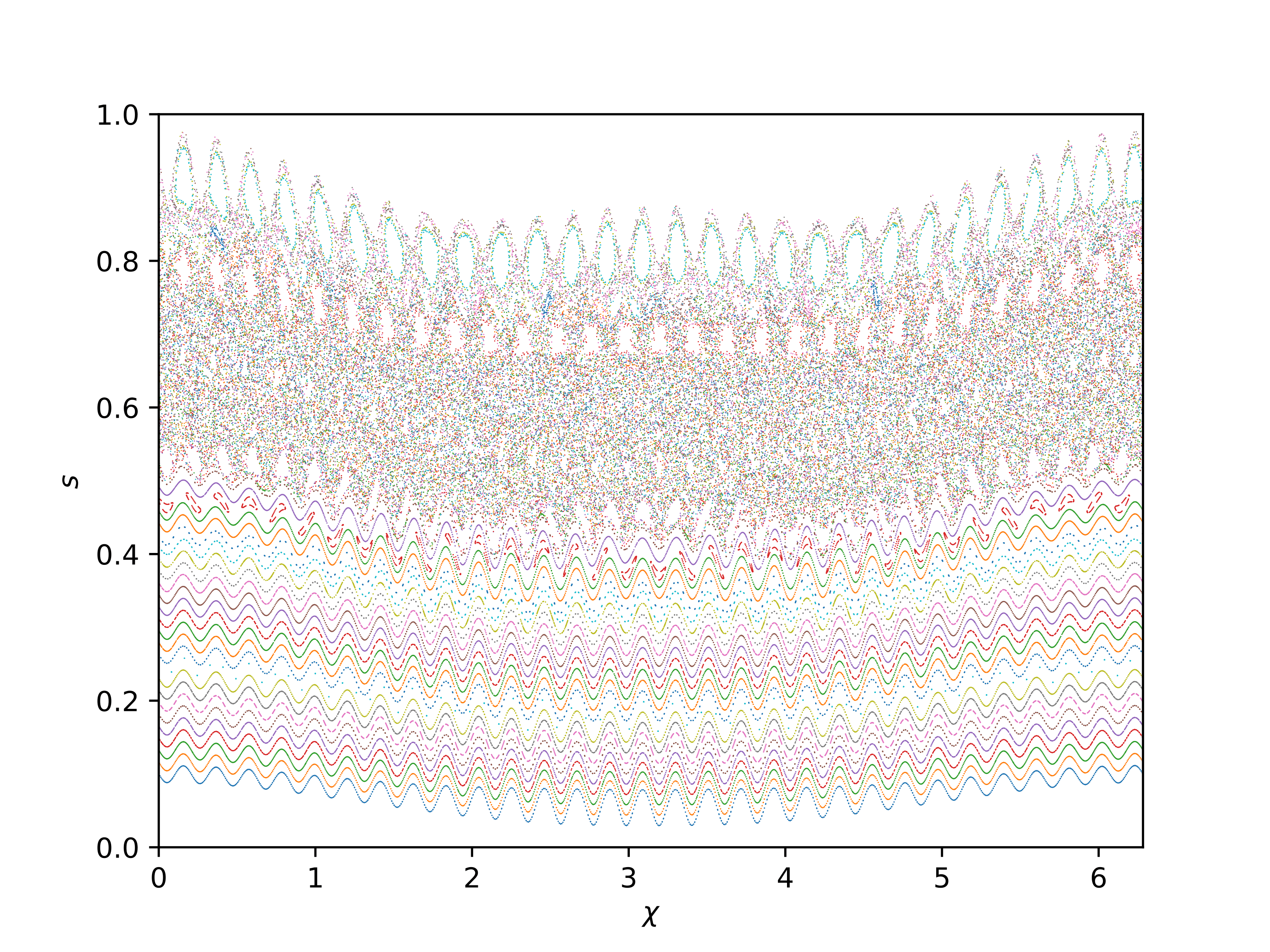}
    \caption{$m = 30$, NCSX}
    \end{subfigure}
    \caption{Kinetic Poincar\'{e} plots are constructed using the mode parameters in Table \ref{tab:mode_parameters} with the perturbation amplitude $\delta \hat{B}^{\psi}= 10^{-3}$. All QS-breaking harmonics of the equilibrium field are artificially suppressed for this analysis.}
    \label{fig:kinetic_poincare_1e-4}
\end{figure}

In Figure \ref{fig:kinetic_poincare_1e-4}, kinetic Poincar\'{e} plots are displayed for particles with $\text{sign}(v_{\|}) = +1$, $\mu = 0$, $E = 3.52$ MeV and Alfv\'{e}nic perturbations with parameters given in Table \ref{tab:mode_parameters}. The amplitude $\hat{\Phi}$ is chosen such that $\delta \hat{B}^{\psi} =  10^{-3}$ on the $s = 1$ surface. Periodic orbits that satisfy the resonance condition $\Omega_l = 0$ appear as $n/(m+l)$ periodic orbits in the kinetic Poincar\'{e} plots.

\begin{figure}
    \centering
    \begin{subfigure}{0.49\textwidth}
    \includegraphics[trim=0.2cm 0.2cm 0.3cm 0.2cm,clip,height=0.7\textwidth]{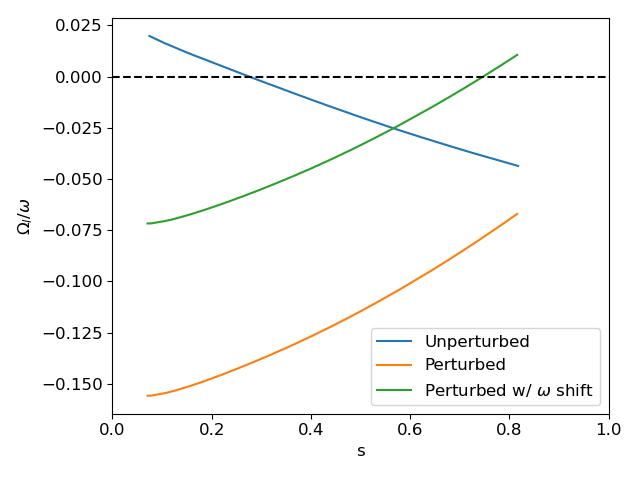}
    \caption{}
    \end{subfigure}
    \begin{subfigure}{0.49\textwidth}
    \includegraphics[trim=0.5cm 0.2cm 1.3cm 8.0cm,clip,height=0.7\textwidth]{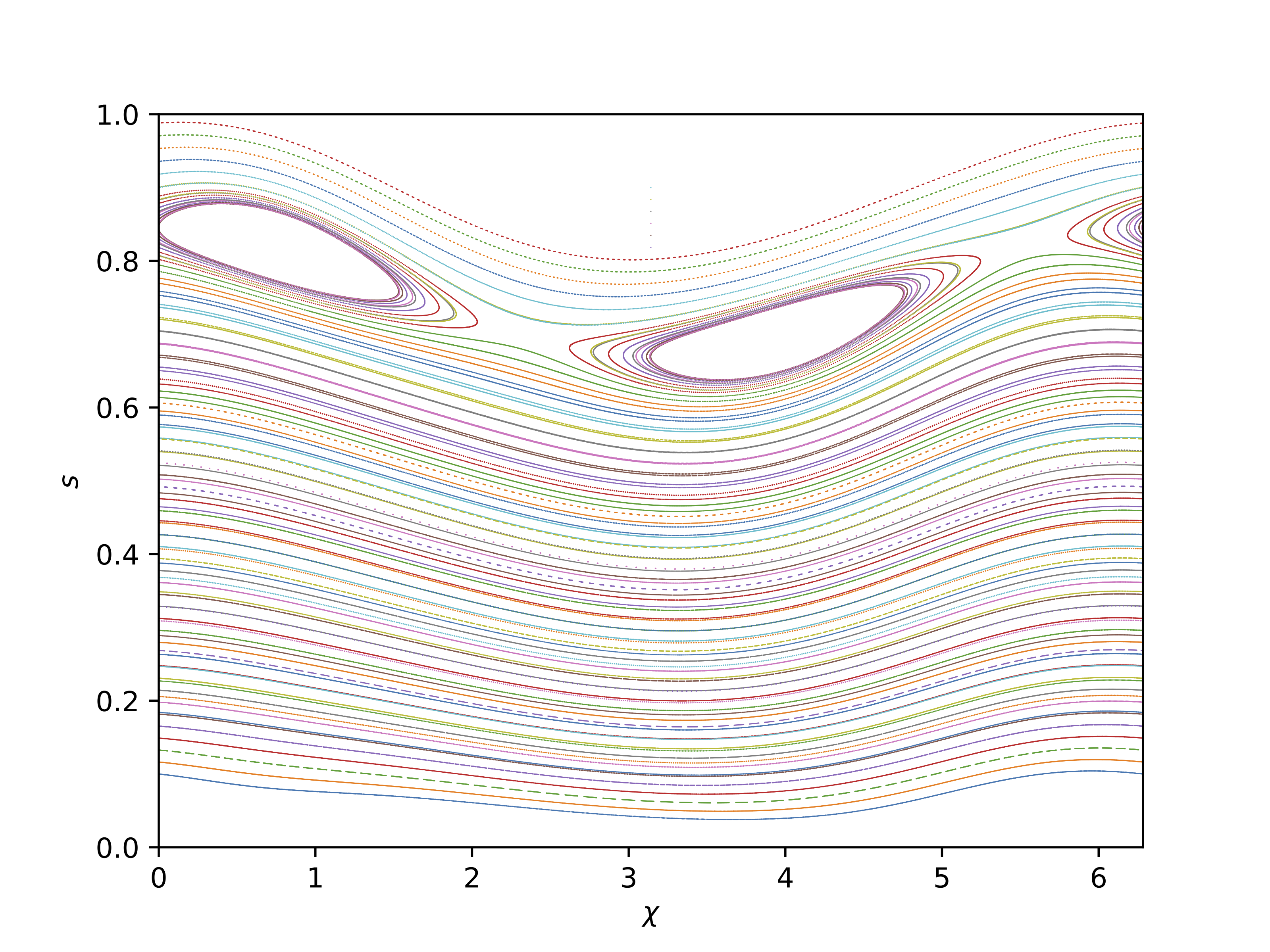}
    \caption{}
    \end{subfigure}
    \caption{(a) In the presence of the resonant perturbation of amplitude $\delta \hat{B}^{\psi} = 10^{-4}$, the characteristic frequencies $\omega_{\theta}$ and $\omega_{\zeta}$ shift, causing the resonance defined by $\Omega_l = 0$ to move outside the equilibrium due to the low shear, see Figure \ref{eq:m_1_vacuum_poinc}. By increasing the mode frequency by 8\%, the resonance reenters. (b) The kinetic Poincar\'{e} plot of amplitude $\delta \hat{B}^{\psi} = 10^{-4}$ with the shifted frequency ($\omega = 2.15$ kHz) reveals the corresponding $l = 1$ island on the shifted resonant surface.}
    \label{fig:QA_vac_shifted}
\end{figure}

For the NCSX and $\beta = 2.5\%$ QA configurations, a clear $l = 1$ island chain is apparent in the presence of the $m = 1$ perturbation. For the vacuum QA equilibrium, the perturbation shifts the orbit helicity slightly, moving the resonance outside the equilibrium due to the low magnetic shear. The resonance reappears by increasing the perturbation frequency by 8\%, see Figure \ref{fig:QA_vac_shifted}. In the $\beta = 2.5\%$ QA equilibria, in addition to the primary $l = 1$ resonance, the $l = 2$ resonance is apparent near $s = 0.1$. However, the island width is small due to the reduced magnitude of the $J_2(\eta_1)$ coupling parameter. As indicated by the resonance plots, Figure \ref{fig:h_0_QA_QH}, none of the other equilibria contain the $m=1$, $l =2$ resonance. 

As the mode number increases from $m = 15$ to $m = 30$, the $l = 1$ island width stays roughly the same for the $\beta = 2.5\%$ QH configuration, as indicated by the scaling of the island width formula \eqref{eq:island_width_basic}-\eqref{eq:psi_cs} for fixed $\delta \hat{B}^{\psi}$. For the vacuum QA configuration, the $l = 1$ resonance reappears for the $m = 15$ and $m = 30$ perturbations, its width being especially wide due to the low magnetic shear of the configuration. For both the vacuum QA and $\beta = 2.5\%$ QH configurations with the $m = 15$ and $m = 30$ perturbations, the island chain is wide enough to lead to visible destruction of nearby KAM surfaces and the formation of secondary island chains. 

In the $\beta = 2.5\%$ QA equilibrium, overlap between the $l = 0$, 1, 2, and 3 resonances is observed with the $m = 15$ perturbation. Substantial island overlap is also observed in the NCSX equilibrium with the $m = 15$ perturbation. In the presence of the $m = 30$ perturbation, strong island overlap is observed in the $\beta = 2.5\%$ QA and NCSX equilibria. However, the phase-space volume over which island overlap occurs is narrowed. As seen in Figure \ref{fig:h_0_QA_QH}, the resonances become more closely spaced for the $m = 30$ perturbation compared to the $m = 15$ perturbation. However, because the island width scales as $\sqrt{\eta_1^{|l|}}$ with $\eta_1 \ll 1$, island overlap does not occur for the large $|l|$ resonant surfaces. This island-width scaling effectively reduces the non-integrable volume as $m$ increases. 

The impact of these phase-space features on the resulting transport will be discussed in Section \ref{sec:Monte_carlo}.

\subsection{Impact of quasisymmetry deviations}

We now investigate the impact of the finite quasisymmetry deviations, quantified in Figure \ref{fig:qs_iota}, on the kinetic Poincar\'{e} analysis. When QS deviations are present, the motion of passing particles becomes 4D $(s,\chi,\zeta,t)$, and the Poincar\'{e} section becomes 3D, $M(s,\chi,\zeta) \rightarrow (s',\chi',\zeta')$. Nonetheless, we can still visualize the Poincar\'{e} map in the $(s,\chi)$ plane to assess the structure of phase space. 

\begin{figure}
    \centering
    \begin{subfigure}{0.49\textwidth}
    \includegraphics[trim=0.5cm 0.2cm 1.3cm 1.2cm,clip,width=1.0\textwidth]{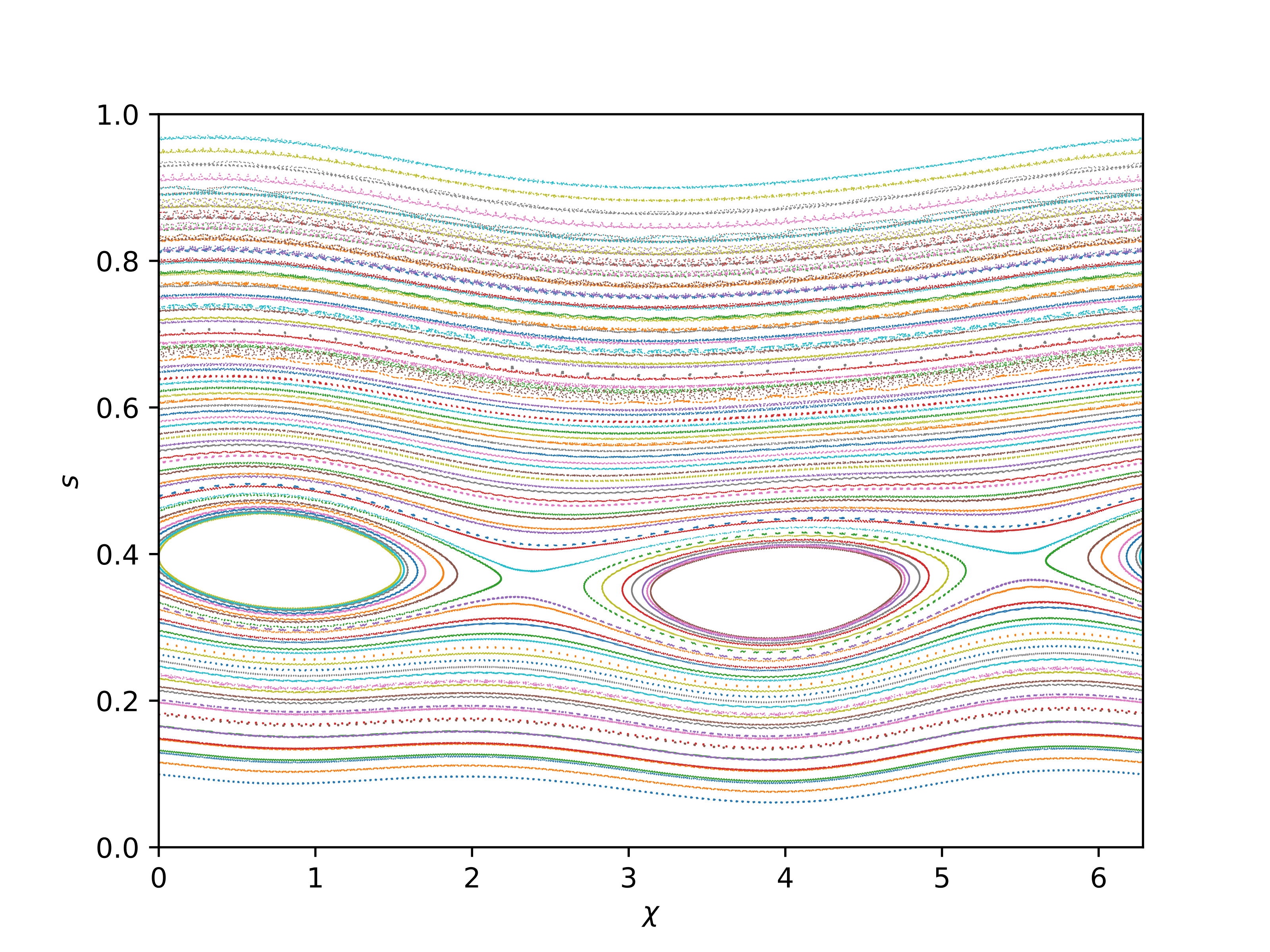}
    \caption{$m = 1$}
    \end{subfigure}
    \begin{subfigure}{0.49\textwidth}
    \includegraphics[trim=0.5cm 0.2cm 1.3cm 1.2cm,clip,width=1.0\textwidth]{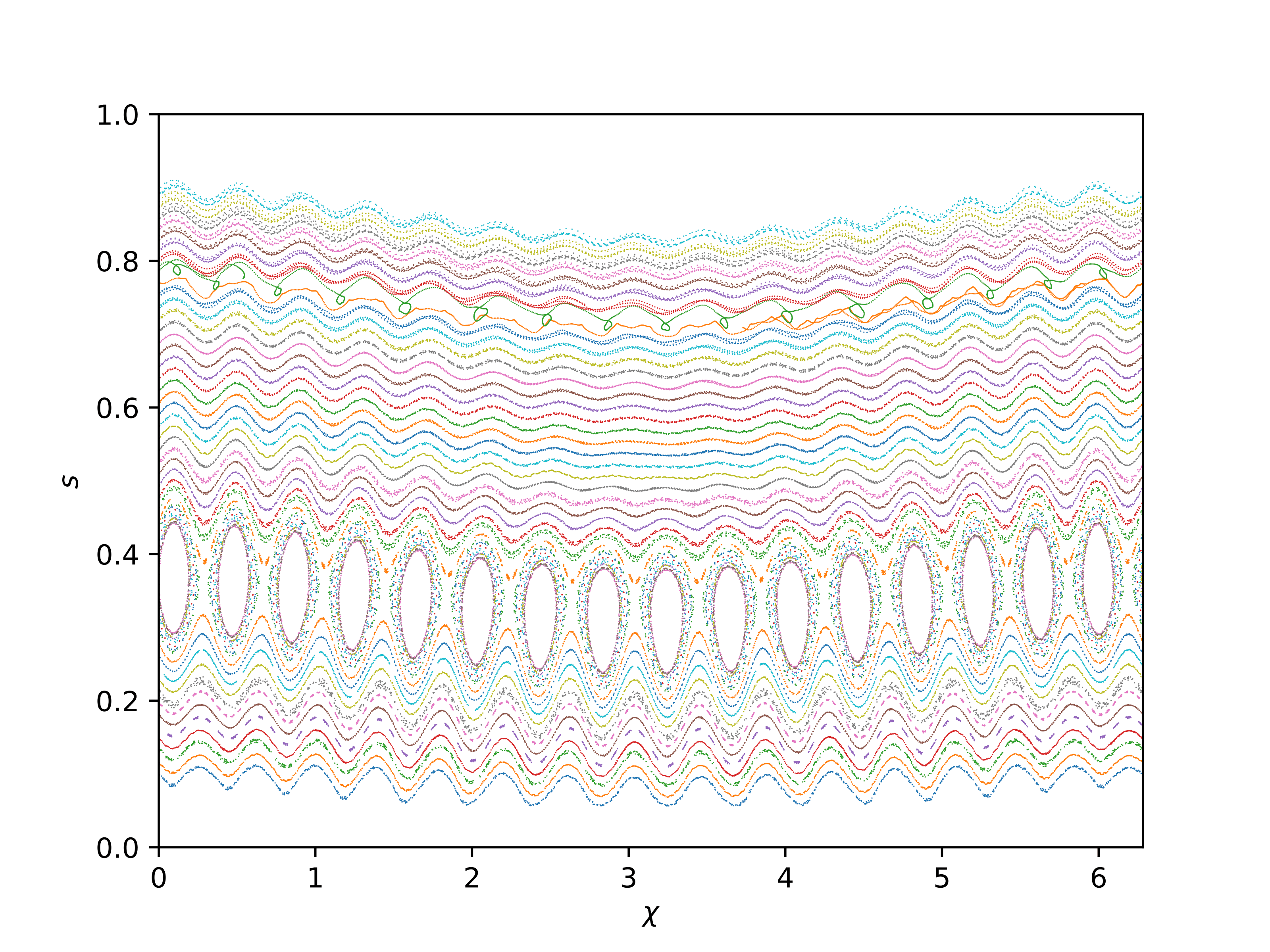}
    \caption{$m = 15$}
    \end{subfigure}
    \caption{$\beta = 2.5\%$ QH kinetic Poincar\'{e} plots with the same parameters as Figure \ref{fig:kinetic_poincare_1e-4}, but without the suppression of the quasisymmetry-breaking modes.}
\label{fig:QH_boots_poincare_perfect_imperfect}
\end{figure}

\begin{figure}
    \centering
    \begin{subfigure}{0.49\textwidth}
    \includegraphics[trim=0.5cm 0.2cm 1.3cm 1.2cm,clip,width=1.0\textwidth]{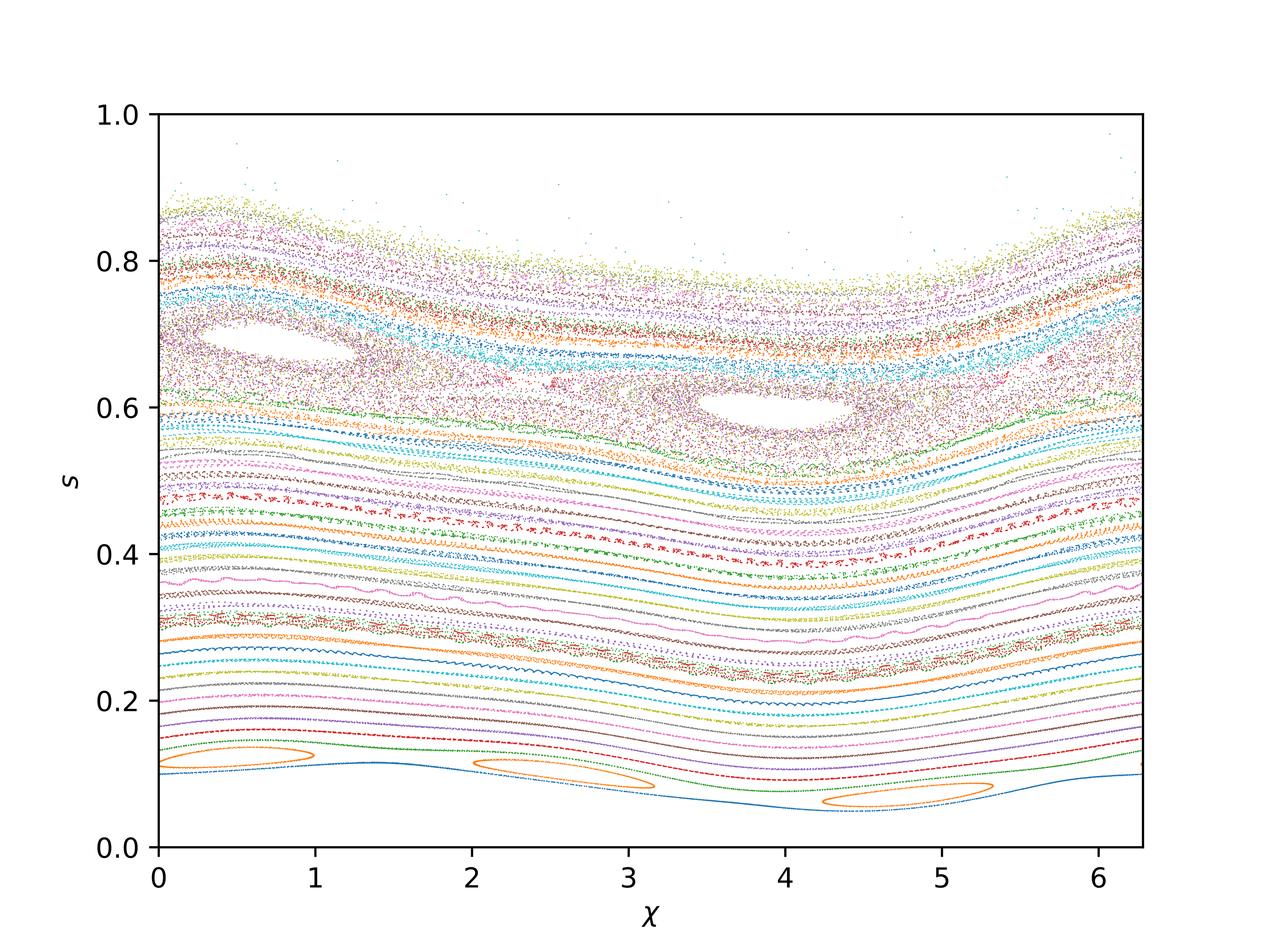}
    \caption{$m = 1$}
    \end{subfigure}
    \begin{subfigure}{0.49\textwidth}
    \includegraphics[trim=0.5cm 0.2cm 1.3cm 1.2cm,clip,width=1.0\textwidth]{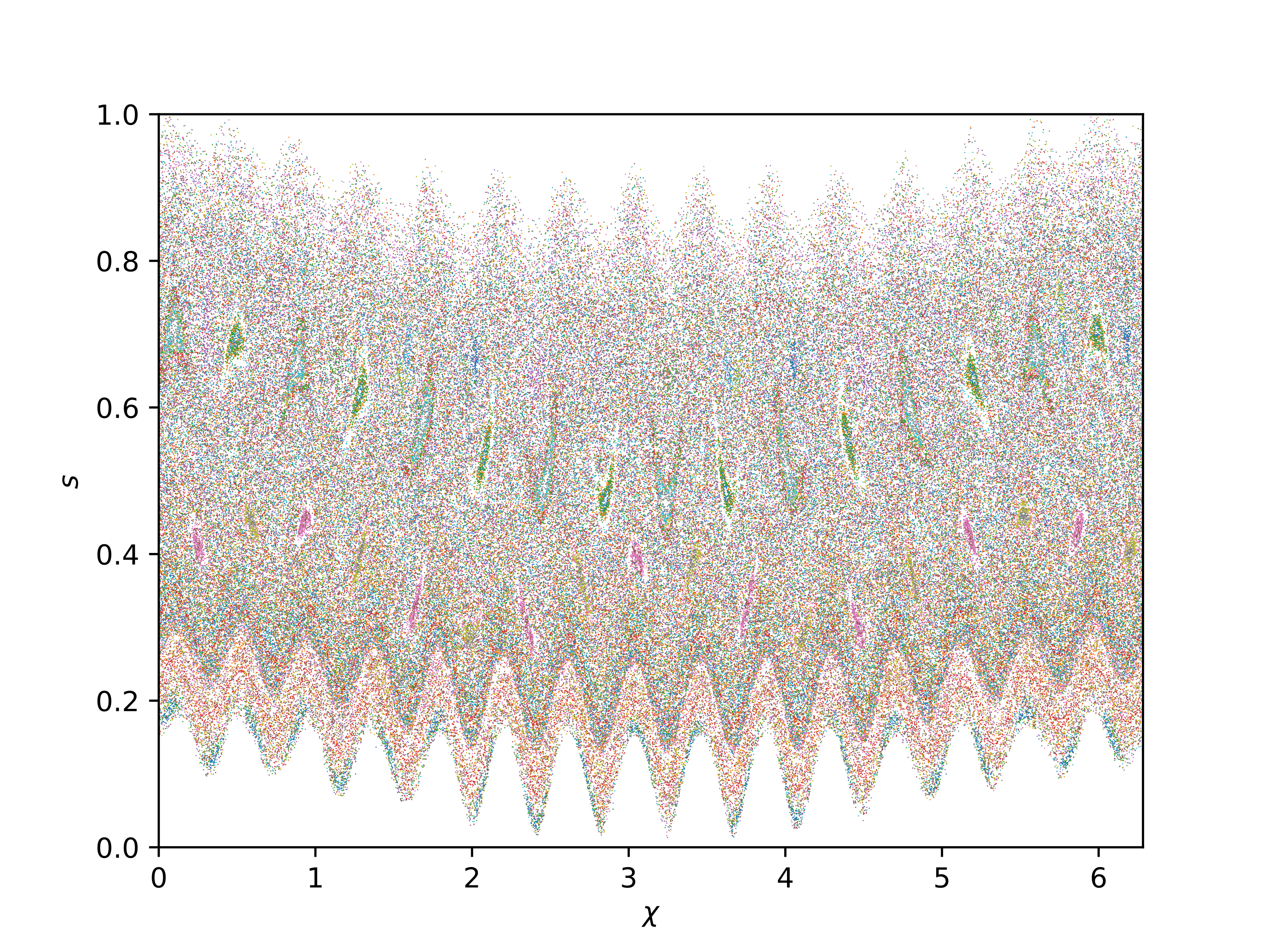}
    \caption{$m = 15$}
    \end{subfigure}
    \caption{NCSX kinetic Poincar\'{e} plots with the same parameters as Figure \ref{fig:kinetic_poincare_1e-4}, but without the suppression of the quasisymmetry-breaking modes.}
    \label{fig:NCSX_poincare_perfect_imperfect}
\end{figure}

In Figures \ref{fig:QH_boots_poincare_perfect_imperfect}-\ref{fig:NCSX_poincare_perfect_imperfect}, we show a selection of the kinetic Poincar\'{e} plots constructed with the same parameters as those in Figure \ref{fig:kinetic_poincare_1e-4}, but without the suppression of the QS-breaking modes. For the case of the $\beta = 2.5\%$ QH equilibrium, the phase-space structure remains mostly unchanged, with the addition of small-scale mixing. In the case of the NCSX equilibrium, the amplitude of the phase-space mixing is amplified due to the enhanced symmetry breaking. The gross phase-space structure remains mostly unchanged in the presence of the $m = 1$ perturbation. In the presence of the $m = 15$ perturbation, the effective diffusion due to the non-integrability of the orbits destroys many of the remaining KAM surfaces and island chains, leading to a wide region of phase-space chaos. We conclude that most large-scale phase space structure is preserved by adding QS deviations, especially for equilibria close to QS, $f_{QS} \sim 10^{-3}$. This finding does not quantitatively agree with recent results \citep{2023White}, indicating substantial diffusive losses in an equilibrium very close to QH ($f_{QS} \sim 10^{-4}$) with an Alfv\'{e}nic perturbation as small as $\delta \hat{B}^{\psi} \sim 10^{-6}$.


\section{Monte Carlo analysis}
\label{sec:Monte_carlo}

\begin{figure}
    \centering 
    \begin{subfigure}{0.49\textwidth}
    \includegraphics[trim=0.2cm 0cm 0.5cm 0.5cm,clip,height=0.75\textwidth]{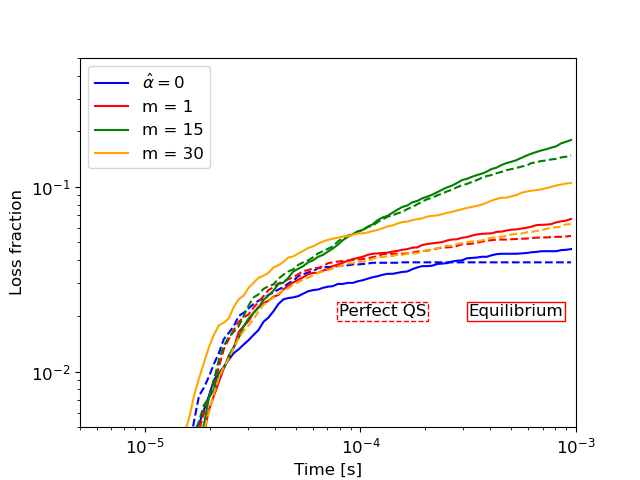}
    \caption{$\beta = 2.5\%$ QA}
    \end{subfigure}
    \begin{subfigure}{0.49\textwidth}
    \includegraphics[trim=0.9cm 0cm 0.5cm 0.5cm,clip,height=0.75\textwidth]{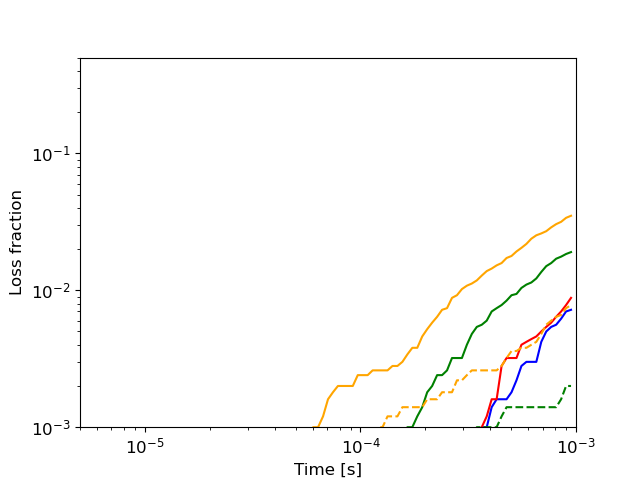}
    \caption{$\beta = 2.5\%$ QH}
    \end{subfigure}
    \begin{subfigure}{0.49\textwidth}
    \includegraphics[trim=0.2cm 0cm 0.5cm 0.5cm,clip,height=0.75\textwidth]{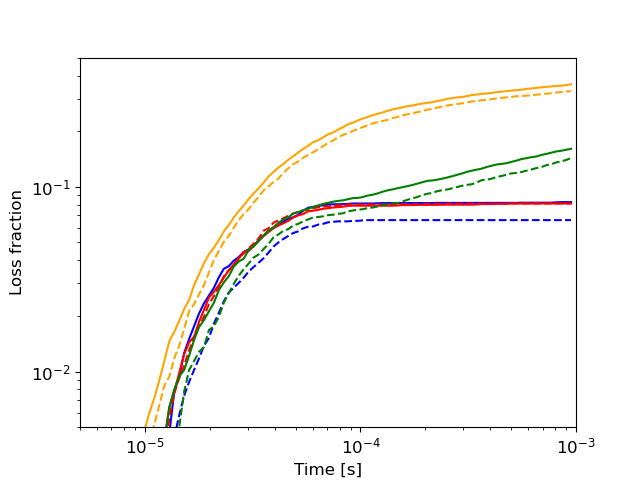}
    \caption{Vacuum QA}
    \end{subfigure} 
    \begin{subfigure}{0.49\textwidth}
    \includegraphics[trim=0.9cm 0cm 0.5cm 0.5cm,clip,height=0.75\textwidth]{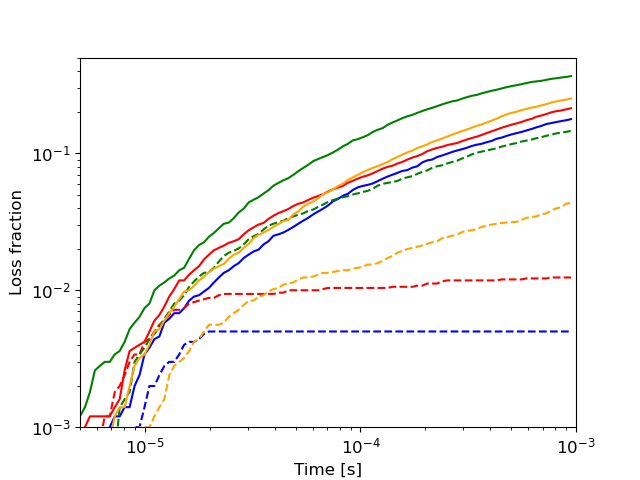}
    \caption{NCSX}
    \end{subfigure}
    \caption{The loss fraction as a function of time is shown. Monte Carlo tracing is performed in the presence of a $\delta \hat{B}^{\psi} = 10^{-3}$ perturbation with parameters described in Table \ref{tab:mode_parameters}. Guiding center trajectories of fusion-born alpha particles are followed for $10^{-3}$ seconds or until they cross through $s=0$ or $s=1$.
    A comparison is made between the actual equilibrium and the equilibrium for which the QS-breaking modes are suppressed, ``perfect QS.''}
    \label{fig:1e-4_loss_history}
\end{figure}

\begin{figure}
    \centering
    \begin{subfigure}{0.49\textwidth}
    \includegraphics[trim=0cm 0cm 1cm 1.0cm,clip,height=0.75\textwidth]{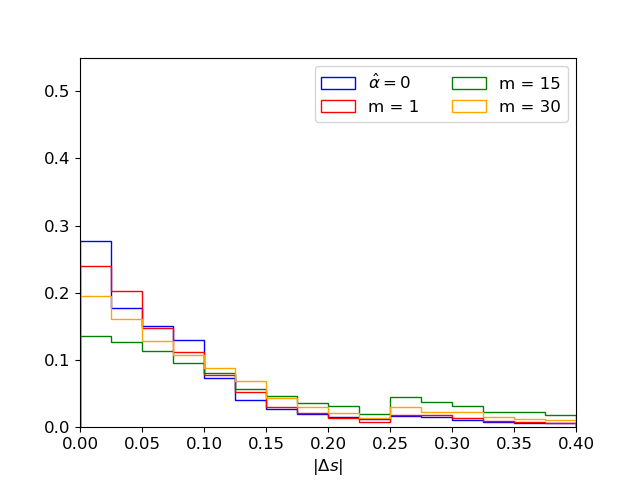}
    \caption{$\beta = 2.5\%$ QA}
    \end{subfigure}
    \begin{subfigure}{0.49\textwidth}
    \includegraphics[trim=0cm 0cm 1cm 1cm,clip,height=0.75\textwidth]{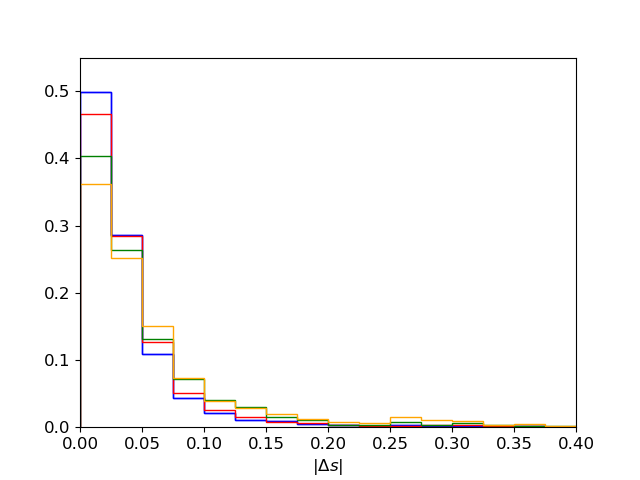}
    \caption{$\beta = 2.5\%$ QH}
    \end{subfigure}
    \begin{subfigure}{0.49\textwidth}
    \includegraphics[trim=0cm 0cm 1cm 1cm,clip,height=0.75\textwidth]{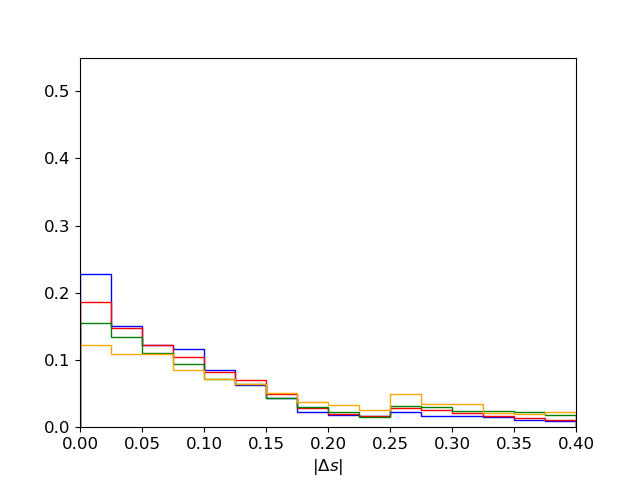}
    \caption{Vacuum QA}
    \end{subfigure}
    \begin{subfigure}{0.49\textwidth}
    \includegraphics[trim=0cm 0cm 1cm 1cm,clip,height=0.75\textwidth]{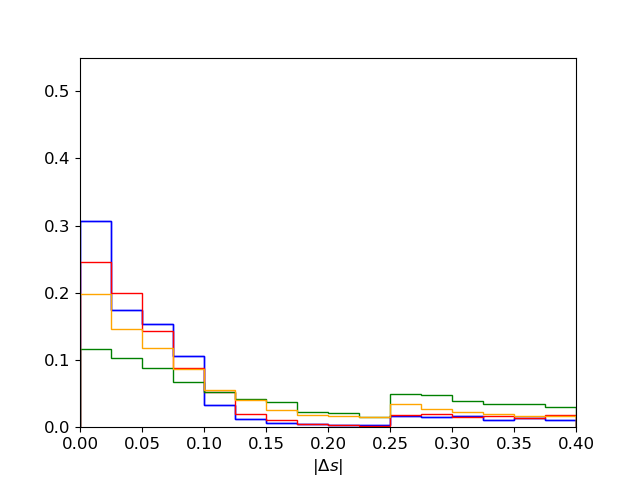}
    \caption{NCSX}
    \end{subfigure}
    \caption{Distribution of radial displacement, $|\Delta s|$, between $t=0$ and the final recorded time  among Monte Carlo samples for the same calculation presented in Figure \ref{fig:1e-4_loss_history}.}
    \label{fig:distributions}
\end{figure}

To assess the impact of the phase-space structure on the resulting transport, we perform Monte Carlo collisionless guiding center tracing simulations. We initialize 5000 particles uniformly in pitch angle and volume for $s \in [0.25, 0.75]$. This initial spatial distribution was chosen to avoid effects related to the coordinate singularity. Particles are followed for $10^{-3}$ seconds or until they are considered lost when they cross through $s = 0$ or $s= 1$.\footnote{Regularity of the equilibrium and perturbed fields near the axis ensures that guiding center trajectories never cross through $s = 0$. Due to the assumption of constant $\delta \hat{\Phi}$, regularity on the axis is not preserved. We, nonetheless, can still gauge the impact of the perturbation on the transport by assessing the loss fraction and flattening of the fast-ion profile.} Figures \ref{fig:1e-4_loss_history}-\ref{fig:distributions} display results of tracing performed with the true equilibria (solid curves, ``equilibrium'') as well as with the quasisymmetry-breaking modes artificially filtered out (dashed curves, ``perfect QS''). Calculations are carried out without Alfv\'{e}nic perturbations ($\hat{\alpha} = 0$) and for the $m = 1$, 15, and 30 perturbations indicated in Table \ref{tab:mode_parameters}.

In Figure \ref{fig:distributions}, the effective transport is quantified by the distribution of the radial displacement, $|\Delta s|$, between $t = 0$ and the final recorded time (after $10^{-3}$ or when a particle is considered lost). By initializing with a uniform distribution over $s$ in a fixed interval, flattening of the distribution function is not as apparent. On the other hand, the distribution of $|\Delta s|$ allows us to identify the net displacement of trajectories due to phase-space islands or chaos.
The double maximum structure reflects the confined trajectories, centered around $|\Delta s| = 0$, and lost trajectories, for which $|\Delta s| \ge 0.25$.
The loss fraction as a function of time is also shown in Figure \ref{fig:1e-4_loss_history}. A summary of the results is presented in Table \ref{tab:Monte_carlo_results}. When comparing the equilibrium and perfect QS cases, we see slight transport enhancement due to QS-breaking modes in the $\beta = 2.5\%$ QA and QH and vacuum QA equilibria. The enhancement of transport becomes much more pronounced in the NCSX equilibrium, given its more substantial deviations from QS; see Figure \ref{fig:qs_iota}. 

In the equilibria that do not exhibit island overlap in Figure \ref{fig:kinetic_poincare_1e-4}, the vacuum QA and $\beta = 2.5\%$ QH equilibria, the total loss fraction increases monotonically with $m$. This result matches the observations in Figure \ref{fig:kinetic_poincare_1e-4}, which indicate that although the island width remains roughly the same, the destruction of nearby KAM surfaces increases with increasing mode number. On the other hand, in equilibria that exhibit strong resonance overlap in Figure \ref{fig:kinetic_poincare_1e-4}, NCSX and $\beta = 2.5\%$ QA, the transport increases monotonically with $m $ until $m = 15$, then decreases for $m = 30$. This characteristic is explained in terms of the kinetic Poincar\'{e} plots in Figure \ref{fig:kinetic_poincare_1e-4}, which indicate that the effective volume of non-integrability is decreased for $m = 30$ compared to $m = 15$ due to the reduction in island width for large $l$ perturbations. 

Overall, the total induced transport is the lowest for the $\beta = 2.5\%$ QH configuration, for which the total losses remain less than $10\%$ in the presence of the Alfv\'{e}nic perturbations. The losses are also less prompt for this configuration, with most beginning around $10^{-4}$ seconds, rather than around $10^{-5}$ seconds for the other configurations. This reduction in transport is due to the increased resonance spacing of quasihelical configurations and its increased magnetic shear compared to the vacuum QA equilibria, see Figure \ref{fig:qs_iota}.

\begin{table}
    \centering
    \begin{tabular}{|c|c|c|c|}
       Configuration & Perturbation & $\text{mean}|\Delta s|$ (Equil./Perfect QS) & Loss frac. (Equil./Perfect QS) \\ \hline 
        $\beta = 2.5\%$ QA & Unperturbed & 0.082/0.085 & 0.046/0.039 \\
        & $m = 1$ & 0.092/0.088 & 0.067/0.054 \\ 
        & $m = 15$ & 0.160/0.143 & 0.179/0.148 \\ 
        & $m = 30$ & 0.114/0.095 & 0.105/0.063 \\ \hline 
        $\beta = 2.5\%$ QH & Unperturbed & 0.037/0.032 & 0.007/0  \\ 
        & $m = 1$ & 0.040/0.033 & 0.009/0  \\ 
        &$m = 15$ & 0.051/0.042 & 0.019/0.002  \\ 
        & $m = 30$ & 0.061/0.046 & 0.035/0.008 \\ \hline 
       NCSX & Unperturbed & 0.127/0.055 & 0.178/0.005 \\ 
        &$m = 1$& 0.144/0.064 & 0.213/0.012 \\ 
        & $m = 15$  & 0.227/0.137 & 0.367/0.146 \\
        & $m = 30$  & 0.170/0.081 & 0.251/0.044 \\ \hline  
        QA vac & Unperturbed & 0.105/0.112 & 0.083/0.066 \\ 
        & $m = 1$ & 0.128/0.132 & 0.125/0.131 \\ 
        & $m = 15$ & 0.163/0.146 & 0.194/0.162  \\ 
        &  $m = 30$ & 0.177/0.167 & 0.203/0.170  \\ 
    \end{tabular}
    \caption{Summary of transport properties for Monte Carlo calculations in the presence of Alfv\'{e}nic perturbations with mode parameters described in Table \ref{tab:mode_parameters} and amplitude $\delta \hat{B}^{\psi} = 10^{-3}$. The mean radial displacement along a trajectory, $\text{mean}|\Delta s|$, and total loss fraction after $10^{-3}$ s are compared between the actual equilibrium, ``equil'', and the equilibrium for which the QS-breaking modes are suppressed, ``perfect QS.''}
    \label{tab:Monte_carlo_results}
\end{table}

\subsection{Transition to phase-space chaos}

\begin{figure}
    \centering
    \begin{subfigure}{0.49\textwidth}
    \includegraphics[trim=0.5cm 0.2cm 1.3cm 1.2cm,clip,width=1.0\textwidth]{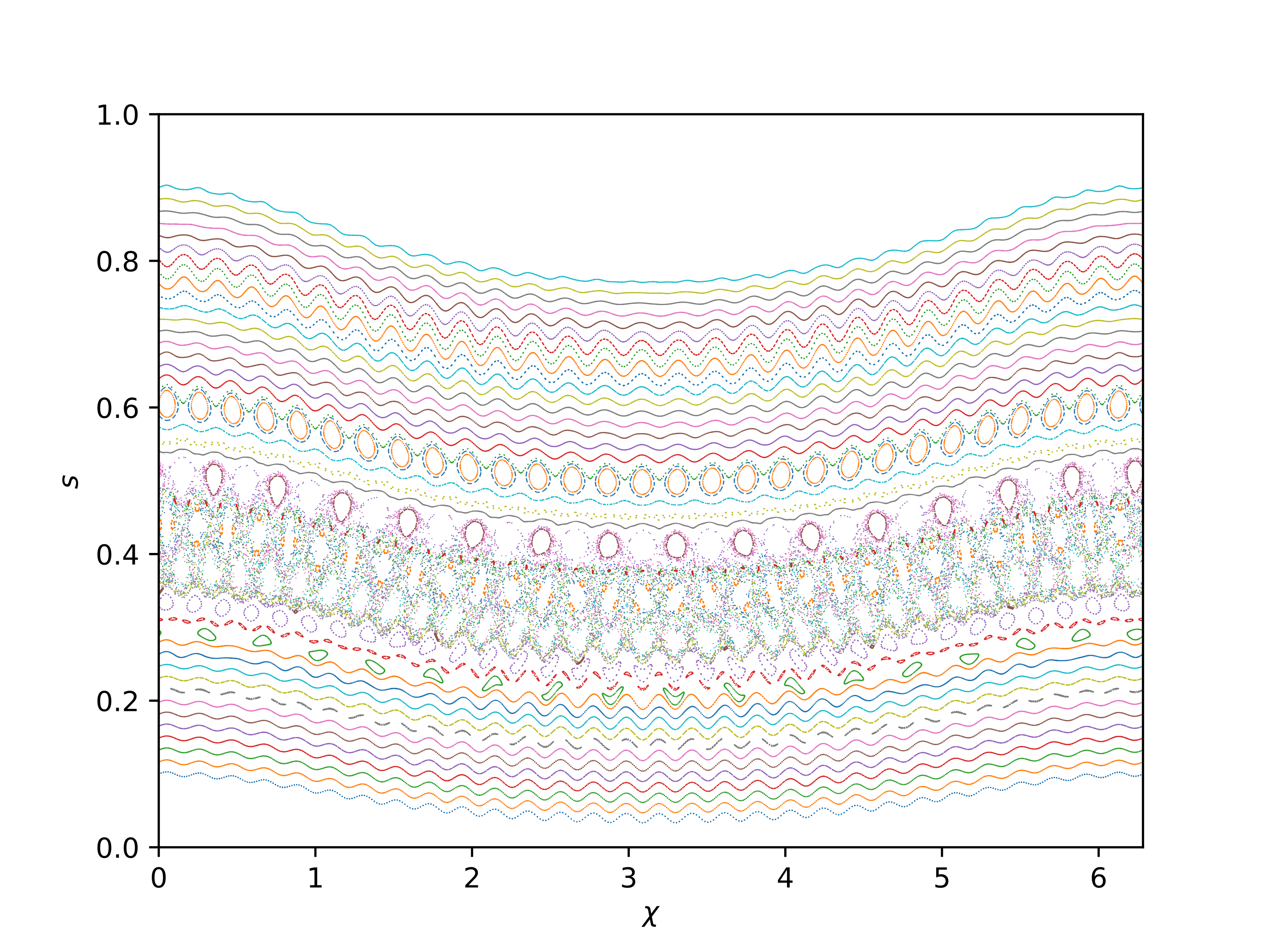}
    \caption{$\delta \hat{B}^{\psi} = 3.14\times 10^{-4}$
    }
    \end{subfigure}
    \begin{subfigure}{0.49\textwidth}
    \includegraphics[trim=0.5cm 0.2cm 1.3cm 1.2cm,clip,width=1.0\textwidth]{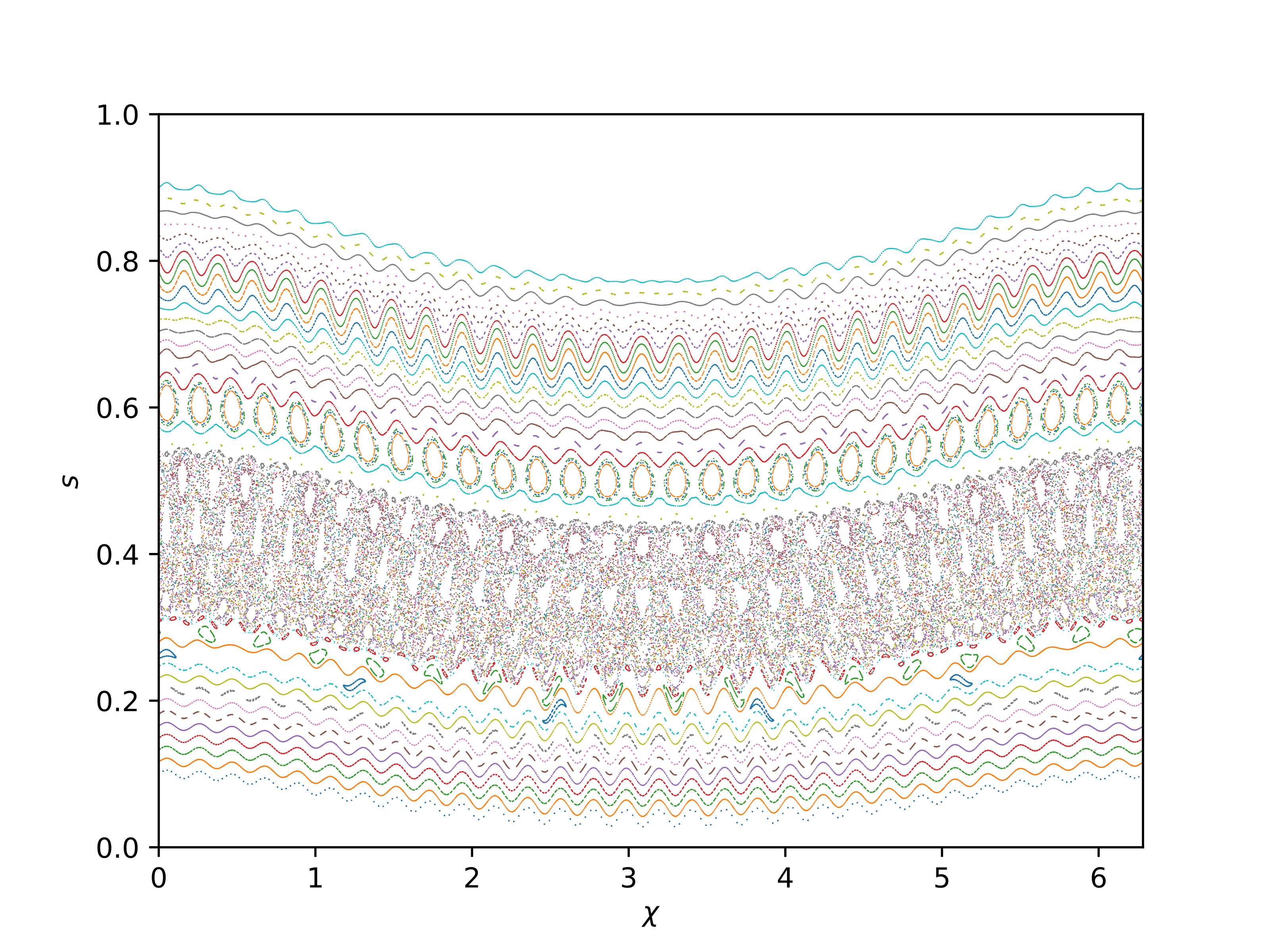}
    \caption{$\delta \hat{B}^{\psi} = 5.58\times 10^{-4}$
    }
    \end{subfigure}
    \begin{subfigure}{0.49\textwidth}
    \includegraphics[trim=0.5cm 0.2cm 1.3cm 1.2cm,clip,width=1.0\textwidth]{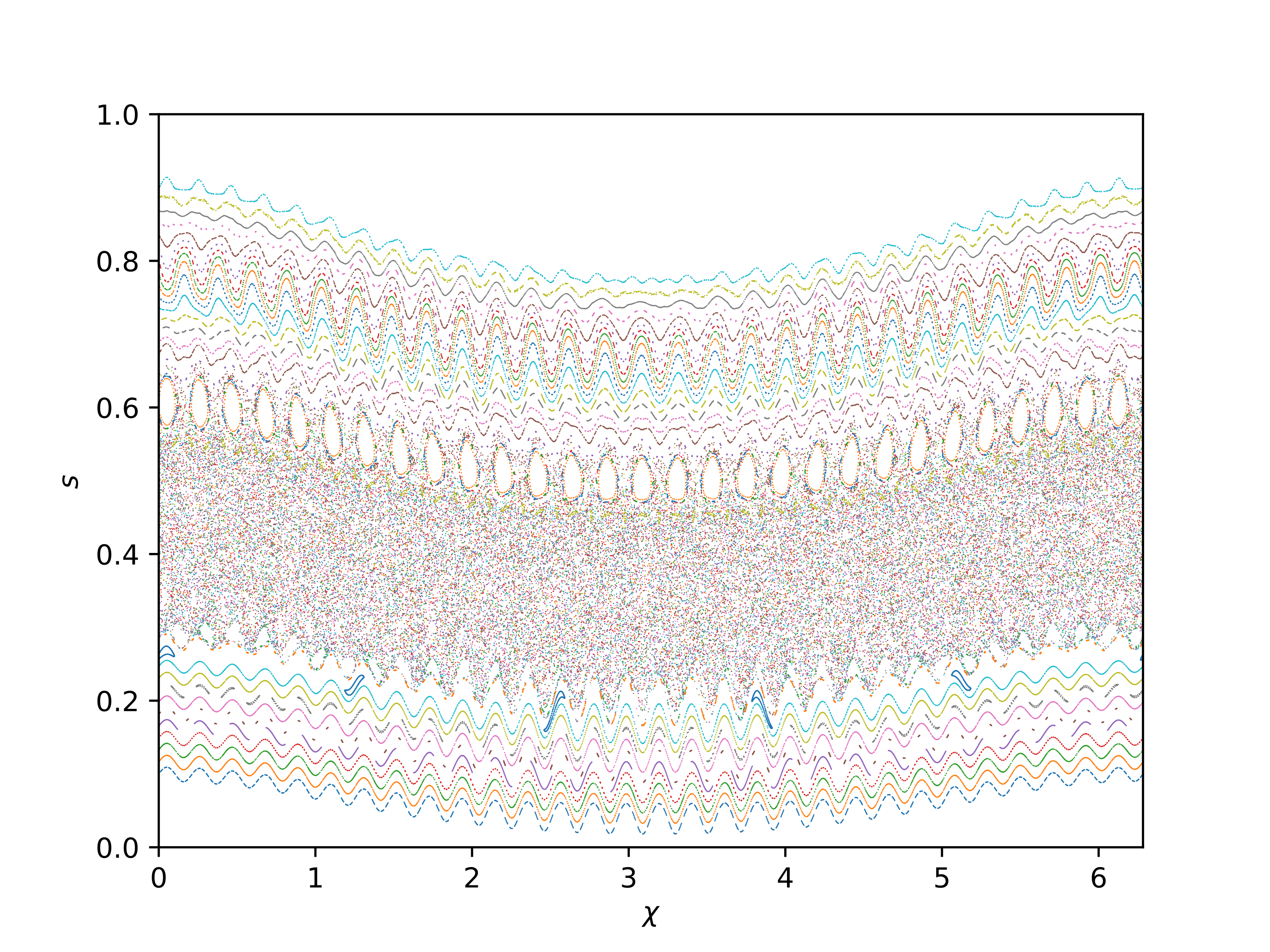}
    \caption{
    $\delta \hat{B}^{\psi} = 9.92\times 10^{-4}$
    }
    \end{subfigure}
    \begin{subfigure}{0.49\textwidth}
    \includegraphics[trim=0.5cm 0.2cm 1.3cm 1.2cm,clip,width=1.0\textwidth]{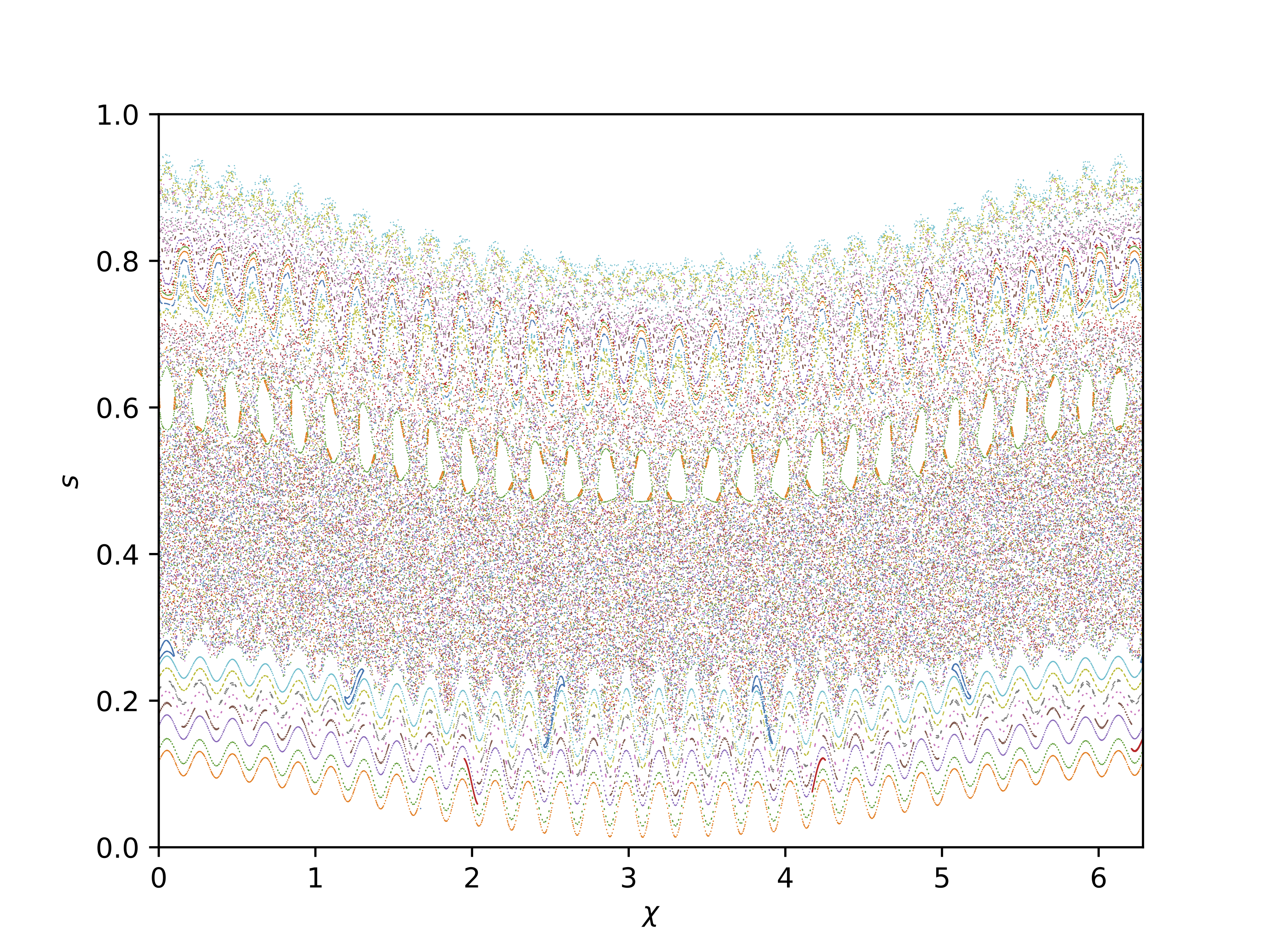}
    \caption{
    $\delta \hat{B}^{\psi} = 1.76\times 10^{-3}$
    }
    \end{subfigure}
    \caption{Kinetic Poincar\'{e} plots for co-passing orbits in the $\beta = 2.5\%$ QA in the presence of an $m = 30$ perturbation with parameters described in Table \ref{tab:mode_parameters}.}
    \label{fig:QA_boots_threshold}
\end{figure}

We now study the transition to phase-space chaos and global transport with increasing perturbation amplitude in the $\beta = 2.5\%$ QA equilibria. We focus on the $m = 30$ perturbation, given that this is the expected mode number at reactor scales. Kinetic Poincar\'{e} plots for co-passing particles resonant with the Alfv\'{e}nic perturbation are shown in Figure \ref{fig:QA_boots_threshold}. The $l = -1$, 0, 1, 2, and 3 islands are visible at the smallest perturbation amplitude, corresponding to $\delta \hat{B}^{\psi} = 3.14 \times 10^{-4}$. As the amplitude is increased to $\delta \hat{B}^{\psi} = 5.58 \times 10^{-4}$, a band of island overlap appears near the resonant surface. The region of destroyed KAM surfaces increases with increasing perturbation amplitude, with only a few remnant islands present at $\delta \hat{B}^{\psi} = 1.76 \times 10^{-3}$. 

\begin{figure}
    \centering 
    \begin{subfigure}{0.49\textwidth}
    \includegraphics[trim=0.2cm 0cm 0.5cm 0.5cm,clip,height=0.75\textwidth]{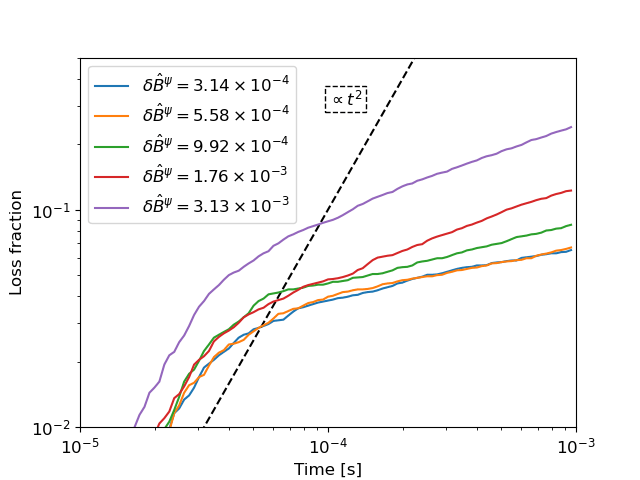}
    \caption{}
    \end{subfigure}
    \begin{subfigure}{0.49\textwidth}
    \includegraphics[trim=0.2cm 0cm 0.5cm 0.5cm,clip,height=0.75\textwidth]{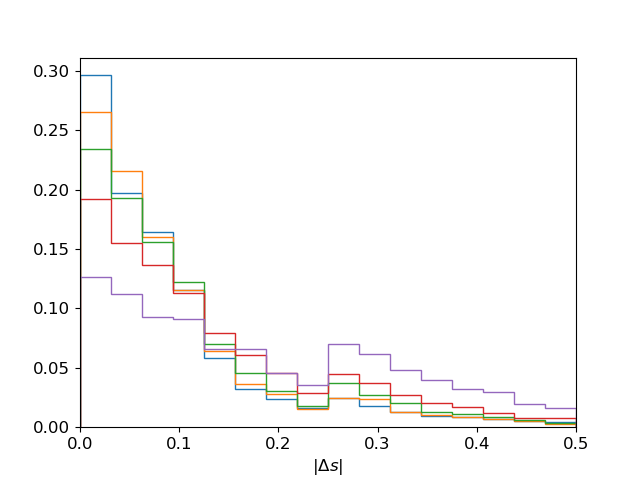}
    \caption{}
    \end{subfigure}
    \begin{subfigure}{0.49\textwidth}
    \includegraphics[trim=0.2cm 0cm 0.5cm 0.5cm,clip,height=0.75\textwidth]{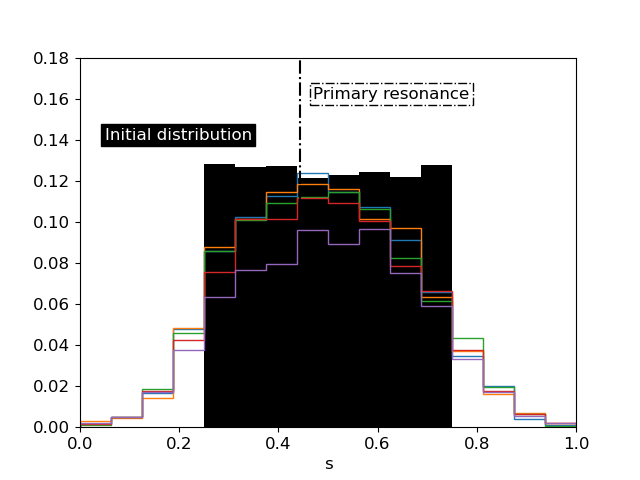}
    \caption{}
    \end{subfigure}
    \begin{subfigure}{0.49\textwidth}
    \includegraphics[trim=0.2cm 0cm 0.5cm 0.5cm,clip,height=0.75\textwidth]{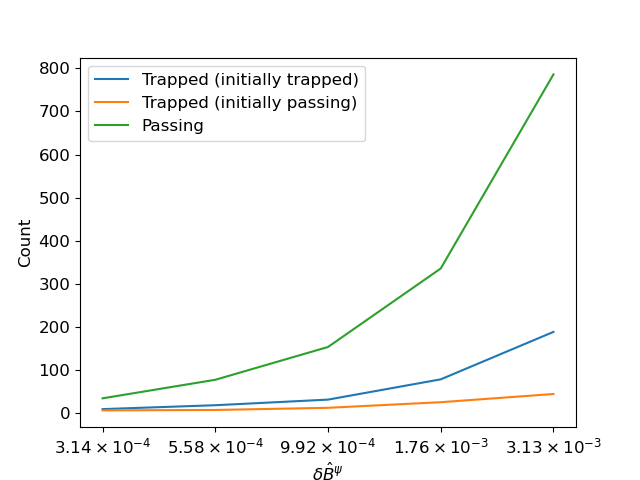}
    \caption{}
    \label{fig:trapped_passing}
    \end{subfigure}
    \caption{Monte Carlo guiding center tracing calculations are performed for the $\beta = 2.5\%$ QA configuration in the presence of an $m = 30$ perturbation with parameters described in Table \ref{tab:mode_parameters}. The perturbation amplitude is increased to study the impact of island overlap, as illustrated in Figure \ref{fig:QA_boots_threshold}, on transport. (a) The loss fraction as a function of time. (b) The distribution of the total radial displacement, $|\Delta s|$, between the initial time and final recorded time. (c) The initial and final radial distribution of particles. (d) The increase in losses of different trajectory classes is plotted as a function of the perturbation amplitudes. The trapped and passing categorization is based on a comparison of the trapping parameter $\lambda = v_{\perp}^2/(v^2B)$ with the maximum field strength on the magnetic surface where a given trajectory is initialized or lost.}
    \label{fig:QA_Monte_carlo}
\end{figure}

Monte Carlo calculations, as described above, are performed for the same Alfv\'{e}nic perturbations to distinguish the impact of the phase-space structure on transport. The loss fraction as a function of time, initial and final distribution function, and net radial displacement $|\Delta s|$ are shown in Figure \ref{fig:QA_Monte_carlo}. We note a sudden increase in the loss fraction above $\delta \hat{B}^{\psi} = 5.58\times 10^{-4}$, the value for which island overlap is seen to occur in Figure \ref{fig:QA_boots_threshold}. This behavior is analogous to the observation of critical gradient behavior on DIII-D, for which the fast-ion transport suddenly becomes stiff above a critical gradient threshold. Similar to our results, the critical gradient behavior on DIII-D arises due to phase-space stochastization \citep{2017Collins}. The initial losses scale approximately as $t^2$, as expected for quasilinear diffusion. As the radial distribution of particles evolves, we see enhanced flattening near the primary resonance surface with increasing perturbation amplitude. 

Figure \ref{fig:trapped_passing} displays the trapping state of lost trajectories as a function of the perturbation amplitude. The trapped and passing status at $t = 0$ and the time of loss are assessed by comparing the trapping parameter $\lambda = v_{\perp}^2/(v^2B)$ with the maximum field strength on the initial and final magnetic surfaces. 
In order to focus on the losses due to the Alfv\'{e}nic perturbation rather than symmetry breaking or wide banana orbits, we subtract the count in each category from the losses without an Alfv\'{e}nic perturbation. As expected, the majority of the loss increase is due to passing orbits. There are additional losses of initially passing orbits that transition to trapped orbits. This loss type can arise from diffusion in the constant of motion space, leading to transformation from a passing orbit to a barely-trapped banana orbit with a large radial width \citep{1992Hsu,1992Sigmar}. Finally, there are losses of trapped particles that do not originate as passing. Further analysis is needed to understand the responsible loss mechanisms. 
Overall, we conclude that the kinetic Poincar\'{e} analysis provides insight into transport characteristics, even for these configurations with finite deviations from quasisymmetry. 

\section{Conclusions}
\label{sec:conclusions}
We have developed the theory for guiding center transport in quasisymmetric equilibria with Alfv\'{e}nic perturbations. Even if the perturbation is restricted to a single $m$ and $n$, additional resonances may be excited due to the coupling of the perturbation to the magnetic drifts, as discussed in Section \ref{sec:resonance_theory}. The resonance condition, phase-space island width, and island overlap conditions are discussed for several equilibria of interest. Quasihelical configurations have a reduced propensity for island overlap due to their increased resonance spacing. While the potential for island overlap increases with the poloidal mode number $m$, the effective volume of phase-space non-integrability decreases when $m$ is large enough due to the reduced drift-island width for higher-order drift couplings. These features are visualized using kinetic Poincar\'{e} plots in Section \ref{sec:kinetic_poincare}. Although the kinetic Poincar\'{e} section is only 2D for a perfectly quasisymmetric equilibrium, this analysis still provides insight into the transport with finite quasisymmetry-breaking errors. The quasisymmetry-breaking errors enhance the transport, especially in configurations with significant unconfined guiding center trajectories such as NCSX. 

We evaluate the transport in several configurations close to quasisymmetry for low ($m = 1$), moderate ($m = 15$), and high ($m = 30$) mode number perturbations, with the highest mode number being that expected for reactor conditions. The toroidal mode number and frequency are chosen to resonate with a drift surface in the equilibrium. We fix the perturbation amplitude to an appropriate amplitude based on experimental measurements and modeling, $\delta B^{r}/B_0 \sim 10^{-3}$, finding that substantial island overlap can be present for moderate and high-mode number perturbations. No island overlap occurs for a quasihelical equilibrium or an equilibrium with very low magnetic shear, such as the recently obtained vacuum equilibria with precise quasisymmetry \citep{2022Landreman}. For the low-shear equilibrium, an enhancement of the loss fraction to $\approx 20\%$ occurs due to the wide orbit width. For the quasihelical configuration, the losses remain less than $7\%$ for all perturbations considered. For configurations with substantial island overlap, the losses increase to $~10-20\%$ due to AE-driven transport. 

Our results are consistent with similar modeling for tokamak configurations \citep{1992Hsu,1992Sigmar}, which indicated resonant overlap and enhanced transport for $\delta B^r/B_0 \sim 10^{-3}$. 
More recent work has highlighted the impact of symmetry-breaking on AE-driven transport in a QH configuration \citep{2023White}. Our results agree qualitatively with their conclusion that unconfined particle orbits enhance diffusive losses. 
However, our results are not in quantitative agreement with their findings of substantial diffusive losses for AE perturbation amplitudes $\delta B^r/B_0 \sim 10^{-6}$ for a QH equilibrium with precise levels of quasisymmetry \citep{2021Landreman}. Further work is necessary in order to resolve this discrepancy. 

As the amplitude of the perturbation is increased, island overlap is observed, leading to a rapid transition to stiff fast ion transport. This result is similar to observations of critical gradient transport in DIII-D \citep{2016Collins,2017Collins} and indicates the potential applicability of quasilinear models \citep{2017Duarte,2018Gorelenkov,2019Duarte} to predict the saturated AE amplitude in stellarators. Since the phase-space island width grows with the amplitude of the magnetic drifts, island overlap is more likely to occur at larger energies. 

Our results suggest several avenues to reduce AE-driven transport in quasisymmetric configurations:
\begin{itemize}
    \item Development of quasihelical configurations, which avoid drift-island overlap due to increased resonance spacing;
    \item Avoiding low magnetic shear, which manifests as wide drift island widths for passing particles;
    \item Evaluating metrics from the unperturbed equilibrium (e.g., the Bessel coupling parameters $J_l(\eta_1)$) as a proxy for drift island width within a stellarator optimization loop.
\end{itemize}

Finally, we remark that many aspects of Alfv\'{e}n eigenmodes were not considered in this study and will affect the transport. To gain a basic understanding of transport, a resonant AE was chosen to have a uniform radial structure. While the radial perturbation amplitude was assumed to be held fixed when comparing across configurations, in practice, the mode structure and amplitude will also depend on the mode numbers and other physical parameters through the AE's nonlinear evolution. 
In the future, we plan to build on this work to obtain a more complete picture of the evolution of AEs in stellarator reactor scenarios. For example, the analysis tools described here, such as kinetic Poincar\'{e} plots, could be used to study the nonlinear evolution of phase-space structures such as hole-clump pairs \citep{2021Bierwage} and zonal structures \citep{2015Zonca} in quasisymmetric configurations.

\section*{Acknowledgements}

The authors would like to acknowledge discussions with Vinicius Duarte and Roscoe White. We acknowledge funding through the U. S. Department of Energy, under contract No. DE-SC0016268, and through the Simons Foundation collaboration ``Hidden
Symmetries and Fusion Energy,'' Grant No.
601958.

\bibliographystyle{jpp}

\bibliography{jpp-instructions}

\begin{thebibliography}{51}
\expandafter\ifx\csname natexlab\endcsname\relax\def\natexlab#1{#1}\fi
\def\au#1{#1} \def\ed#1{#1} \def\yr#1{#1}\def\at#1{#1}\def\jt#1{\textit{#1}} \def\bt#1{#1}\def\bvol#1{\textbf{#1}} \def\vol#1{#1} \def\pg#1{#1} \def\publ#1{#1}\def\arxiv#1{#1}\def\org#1{#1}\def\st#1{\textit{#1}}

\bibitem[Bader {\em et~al.\/}(2021)Bader, Anderson, Drevlak, Faber, Hegna, Henneberg, Landreman, Schmitt, Suzuki \& Ware]{2021Bader}
{\sc \au{Bader, A.}, \au{Anderson, D.~T.}, \au{Drevlak, M.}, \au{Faber, B.~J.}, \au{Hegna, C.~C.}, \au{Henneberg, S.}, \au{Landreman, M.}, \au{Schmitt, J.~C.}, \au{Suzuki, Y.} \& \au{Ware, A.}} \yr{2021}  \at{Modeling of energetic particle transport in optimized stellarators}.  \jt{Nuclear Fusion}  \bvol{61}~(11),  \pg{116060}.

\bibitem[Bader {\em et~al.\/}(2019)Bader, Drevlak, Anderson, Faber, Hegna, Likin, Schmitt \& Talmadge]{2019Bader}
{\sc \au{Bader, A.}, \au{Drevlak, M.}, \au{Anderson, D.~T.}, \au{Faber, B.~J.}, \au{Hegna, C.~C.}, \au{Likin, K.~M.}, \au{Schmitt, J.~C.} \& \au{Talmadge, J.~N.}} \yr{2019}  \at{Stellarator equilibria with reactor relevant energetic particle losses}.  \jt{Journal of Plasma Physics}  \bvol{85}~(5).

\bibitem[Bierwage {\em et~al.\/}(2021)Bierwage, White \& Duarte]{2021Bierwage}
{\sc \au{Bierwage, A.}, \au{White, R.~B.} \& \au{Duarte, V.~N.}} \yr{2021}  \at{On the effect of beating during nonlinear frequency chirping}.  \jt{Plasma and Fusion Research}  \bvol{16},  \pg{1403087--1403087}.

\bibitem[Collins {\em et~al.\/}(2016)Collins, Heidbrink, Austin, Kramer, Pace, Petty, Stagner, Van~Zeeland, White, Zhu {\em et~al.\/}]{2016Collins}
{\sc \au{Collins, C.~S.}, \au{Heidbrink, W.~W.}, \au{Austin, M.~E.}, \au{Kramer, G.~J.}, \au{Pace, D.~C.}, \au{Petty, C.~C.}, \au{Stagner, L.}, \au{Van~Zeeland, M.~A.}, \au{White, R.~B.}, \au{Zhu, Y.~B.} \& \au{others}} \yr{2016}  \at{{Observation of critical-gradient behavior in {Alfv{\'e}n}-eigenmode-induced fast-ion transport}}.  \jt{Physical Review Letters}  \bvol{116}~(9),  \pg{095001}.

\bibitem[Collins {\em et~al.\/}(2017)Collins, Heidbrink, Podest{\`a}, White, Kramer, Pace, Petty, Stagner, Van~Zeeland, Zhu {\em et~al.\/}]{2017Collins}
{\sc \au{Collins, C.~S.}, \au{Heidbrink, W.~W.}, \au{Podest{\`a}, M.}, \au{White, R.~B.}, \au{Kramer, G.~J.}, \au{Pace, D.~C.}, \au{Petty, C.~C.}, \au{Stagner, L.}, \au{Van~Zeeland, M.~A.}, \au{Zhu, Y.~B.} \& \au{others}} \yr{2017}  \at{Phase-space dependent critical gradient behavior of fast-ion transport due to {Alfv{\'e}n} eigenmodes}.  \jt{Nuclear Fusion}  \bvol{57}~(8),  \pg{086005}.

\bibitem[Crocker {\em et~al.\/}(2013)Crocker, Fredrickson, Gorelenkov, Peebles, Kubota, Bell, Diallo, LeBlanc, Menard, Podesta {\em et~al.\/}]{2013Crocker}
{\sc \au{Crocker, N.~A.}, \au{Fredrickson, E.~D.}, \au{Gorelenkov, N.~N.}, \au{Peebles, W.~A.}, \au{Kubota, S.}, \au{Bell, R.~E.}, \au{Diallo, A.}, \au{LeBlanc, B.~P.}, \au{Menard, J.~E.}, \au{Podesta, M.} \& \au{others}} \yr{2013}  \at{Internal amplitude, structure and identification of compressional and global {Alfv{\'e}n} eigenmodes in {NSTX}}.  \jt{Nuclear Fusion}  \bvol{53}~(4),  \pg{043017}.

\bibitem[Das {\em et~al.\/}(2016)Das, Saiki, Sander \& Yorke]{2016Das}
{\sc \au{Das, S.}, \au{Saiki, Y.}, \au{Sander, E.} \& \au{Yorke, J.~A.}} \yr{2016}  \at{Quasiperiodicity: rotation numbers}.  \bt{In {\em The Foundations of Chaos Revisited: From Poincar{\'e} to Recent Advancements\/}},  \pg{pp. 103--118}.  \publ{Springer}.

\bibitem[Deng {\em et~al.\/}(2009)Deng, Brower, Breizman, Spong, Almagri, Anderson, Anderson, Ding, Guttenfelder, Likin {\em et~al.\/}]{2009Deng}
{\sc \au{Deng, C.~B.}, \au{Brower, D.~L.}, \au{Breizman, B.~N.}, \au{Spong, D.~A.}, \au{Almagri, A.~F.}, \au{Anderson, D.~T.}, \au{Anderson, F. S.~B.}, \au{Ding, W.~X.}, \au{Guttenfelder, W.}, \au{Likin, K.~M.} \& \au{others}} \yr{2009}  \at{Energetic-electron-driven instability in the {Helically Symmetric Experiment}}.  \jt{Physical Review Letters}  \bvol{103}~(2),  \pg{025003}.

\bibitem[Duarte {\em et~al.\/}(2017)Duarte, Berk, Gorelenkov, Heidbrink, Kramer, Nazikian, Pace, Podesta, Tobias \& Van~Zeeland]{2017Duarte}
{\sc \au{Duarte, V.~N.}, \au{Berk, H.~L.}, \au{Gorelenkov, N.~N.}, \au{Heidbrink, W.~W.}, \au{Kramer, G.~J.}, \au{Nazikian, R.}, \au{Pace, D.~C.}, \au{Podesta, M.}, \au{Tobias, B.~J.} \& \au{Van~Zeeland, M.~A.}} \yr{2017}  \at{Prediction of nonlinear evolution character of energetic-particle-driven instabilities}.  \jt{Nuclear Fusion}  \bvol{57}~(5),  \pg{054001}.

\bibitem[Duarte {\em et~al.\/}(2019)Duarte, Gorelenkov, White \& Berk]{2019Duarte}
{\sc \au{Duarte, V.~N.}, \au{Gorelenkov, N.~N.}, \au{White, R.~B.} \& \au{Berk, H.~L.}} \yr{2019}  \at{Collisional resonance function in discrete-resonance quasilinear plasma systems}.  \jt{Physics of Plasmas}  \bvol{26}~(12),  \pg{120701}.

\bibitem[Feh{\'e}r(2014)]{2014Feher}
{\sc \au{Feh{\'e}r, T.~B.}} \yr{2014}  \at{Simulation of the interaction between {Alfv{\'e}n} waves and fast particles}. PhD thesis, University of Greifswald.

\bibitem[Garren \& Boozer(1991)]{1991Garren}
{\sc \au{Garren, D.~A.} \& \au{Boozer, A.~H.}} \yr{1991}  \at{Magnetic field strength of toroidal plasma equilibria}.  \jt{Physics of Fluids B: Plasma Physics}  \bvol{3}~(10),  \pg{2805--2821}.

\bibitem[Gorelenkov {\em et~al.\/}(2018)Gorelenkov, Duarte, Podesta \& Berk]{2018Gorelenkov}
{\sc \au{Gorelenkov, N.~N.}, \au{Duarte, V.~N.}, \au{Podesta, M.} \& \au{Berk, H.~L.}} \yr{2018}  \at{{Resonance broadened quasi-linear (RBQ) model for fast ion distribution relaxation due to Alfv{\'e}nic eigenmodes}}.  \jt{Nuclear Fusion}  \bvol{58}~(8),  \pg{082016}.

\bibitem[Gorelenkov {\em et~al.\/}(2014)Gorelenkov, Pinches \& Toi]{2014Gorelenkov}
{\sc \au{Gorelenkov, N.~N.}, \au{Pinches, S.~D.} \& \au{Toi, K.}} \yr{2014}  \at{Energetic particle physics in fusion research in preparation for burning plasma experiments}.  \jt{Nuclear Fusion}  \bvol{54}~(12),  \pg{125001}.

\bibitem[Heidbrink {\em et~al.\/}(2008)Heidbrink, Van~Zeeland, Austin, Burrell, Gorelenkov, Kramer, Luo, Makowski, McKee, Muscatello {\em et~al.\/}]{2008Heidbrink}
{\sc \au{Heidbrink, W.~W.}, \au{Van~Zeeland, M.~A.}, \au{Austin, M.~E.}, \au{Burrell, K.~H.}, \au{Gorelenkov, N.~N.}, \au{Kramer, G.~J.}, \au{Luo, Y.}, \au{Makowski, M.~A.}, \au{McKee, G.~R.}, \au{Muscatello, C.} \& \au{others}} \yr{2008}  \at{Central flattening of the fast-ion profile in reversed-shear {DIII-D} discharges}.  \jt{Nuclear Fusion}  \bvol{48}~(8),  \pg{084001}.

\bibitem[Helander {\em et~al.\/}(2012)Helander, Beidler, Bird, Drevlak, Feng, Hatzky, Jenko, Kleiber, Proll, Turkin {\em et~al.\/}]{2012Helander}
{\sc \au{Helander, P.}, \au{Beidler, C.~D.}, \au{Bird, T.~M.}, \au{Drevlak, M.}, \au{Feng, Y.}, \au{Hatzky, R.}, \au{Jenko, F.}, \au{Kleiber, R.}, \au{Proll, J. H.~E.}, \au{Turkin, Y.} \& \au{others}} \yr{2012}  \at{Stellarator and tokamak plasmas: a comparison}.  \jt{Plasma Physics and Controlled Fusion}  \bvol{54}~(12),  \pg{124009}.

\bibitem[Hsu \& Sigmar(1992)]{1992Hsu}
{\sc \au{Hsu, C.-T.} \& \au{Sigmar, D.~J.}} \yr{1992}  \at{{Alpha-particle losses from toroidicity-induced {Alfv{\'e}n eigenmodes. Part I: Phase-space topology of energetic particle orbits in tokamak plasma}}}.  \jt{Physics of Fluids B: Plasma Physics}  \bvol{4}~(6),  \pg{1492--1505}.

\bibitem[Koniges {\em et~al.\/}(2003)Koniges, Grossman, Fenstermacher, Kisslinger, Mioduszewski, Rognlien, Strumberger \& Umansky]{2003Koniges}
{\sc \au{Koniges, A.~E.}, \au{Grossman, A.}, \au{Fenstermacher, M.}, \au{Kisslinger, J.}, \au{Mioduszewski, P.}, \au{Rognlien, T.}, \au{Strumberger, E.} \& \au{Umansky, M.}} \yr{2003}  \at{Magnetic topology of a candidate {NCSX} plasma boundary configuration}.  \jt{Nuclear Fusion}  \bvol{43}~(2),  \pg{107}.

\bibitem[Kruger {\em et~al.\/}(1998)Kruger, Hegna \& Callen]{1998Kruger}
{\sc \au{Kruger, S.~E.}, \au{Hegna, C.~C.} \& \au{Callen, J.~D.}} \yr{1998}  \at{Generalized reduced magnetohydrodynamic equations}.  \jt{Physics of Plasmas}  \bvol{5}~(12),  \pg{4169--4182}.

\bibitem[Ku {\em et~al.\/}(2008)Ku, Garabedian, Lyon, Turnbull, Grossman, Mau, Zarnstorff \& {ARIES Team}]{2008Ku}
{\sc \au{Ku, L.~P.}, \au{Garabedian, P.~R.}, \au{Lyon, J.}, \au{Turnbull, A.}, \au{Grossman, A.}, \au{Mau, T.~K.}, \au{Zarnstorff, M.} \& \au{{ARIES Team}}} \yr{2008}  \at{Physics design for {ARIES-CS}}.  \jt{Fusion Science and Technology}  \bvol{54}~(3),  \pg{673--693}.

\bibitem[Landreman {\em et~al.\/}(2022)Landreman, Buller \& Drevlak]{2022Landremanb}
{\sc \au{Landreman, M.}, \au{Buller, S.} \& \au{Drevlak, M.}} \yr{2022}  \at{Optimization of quasi-symmetric stellarators with self-consistent bootstrap current and energetic particle confinement}.  \jt{Physics of Plasmas}  \bvol{29}~(8),  \pg{082501}.

\bibitem[Landreman {\em et~al.\/}(2021)Landreman, Medasani, Wechsung, Giuliani, Jorge \& Zhu]{2021Landreman}
{\sc \au{Landreman, M.}, \au{Medasani, B.}, \au{Wechsung, F.}, \au{Giuliani, A.}, \au{Jorge, R.} \& \au{Zhu, C.}} \yr{2021}  \at{{SIMSOPT: A flexible framework for stellarator optimization}}.  \jt{Journal of Open Source Software}  \bvol{6}~(65),  \pg{3525}.

\bibitem[Landreman \& Paul(2022)]{2022Landreman}
{\sc \au{Landreman, M.} \& \au{Paul, E.~J.}} \yr{2022}  \at{Magnetic fields with precise quasisymmetry for plasma confinement}.  \jt{Physical Review Letters}  \bvol{128}~(3),  \pg{035001}.

\bibitem[LeViness {\em et~al.\/}(2022)LeViness, Schmitt, Lazerson, Bader, Faber, Hammond \& Gates]{2022Leviness}
{\sc \au{LeViness, A.}, \au{Schmitt, J.~C.}, \au{Lazerson, S.~A.}, \au{Bader, A.}, \au{Faber, B.~J.}, \au{Hammond, K.~C.} \& \au{Gates, D.~A.}} \yr{2022}  \at{Energetic particle optimization of quasi-axisymmetric stellarator equilibria}.  \jt{Nuclear Fusion}  \bvol{63}~(1),  \pg{016018}.

\bibitem[Littlejohn(1983)]{1983Littlejohn}
{\sc \au{Littlejohn, R.~G.}} \yr{1983}  \at{Variational principles of guiding centre motion}.  \jt{Journal of Plasma Physics}  \bvol{29}~(1),  \pg{111--125}.

\bibitem[Littlejohn(1985)]{1985Littlejohn}
{\sc \au{Littlejohn, R.~G.}} \yr{1985}  \at{Differential forms and canonical variables for drift motion in toroidal geometry}.  \jt{The Physics of Fluids}  \bvol{28}~(6),  \pg{2015--2016}.

\bibitem[Melnikov {\em et~al.\/}(2014)Melnikov, Ochando, Ascasibar, Castejon, Cappa, Eliseev, Hidalgo, Krupnik, Lopez-Fraguas, Liniers {\em et~al.\/}]{2014Melnikov}
{\sc \au{Melnikov, A.~V.}, \au{Ochando, M.}, \au{Ascasibar, E.}, \au{Castejon, F.}, \au{Cappa, A.}, \au{Eliseev, L.~G.}, \au{Hidalgo, C.}, \au{Krupnik, L.~I.}, \au{Lopez-Fraguas, A.}, \au{Liniers, M.} \& \au{others}} \yr{2014}  \at{{Effect of magnetic configuration on frequency of NBI-driven Alfv{\'e}n modes in TJ-II}}.  \jt{Nuclear Fusion}  \bvol{54}~(12),  \pg{123002}.

\bibitem[Mynick(1993{\natexlab{{\em a\/}}})]{1993Mynickb}
{\sc \au{Mynick, H.~E.}} \yr{1993{\natexlab{{\em a\/}}}}  \at{Stochastic transport of {MeV} ions by low-n magnetic perturbations}.  \jt{Physics of Fluids B: Plasma Physics}  \bvol{5}~(7),  \pg{2460--2467}.

\bibitem[Mynick(1993{\natexlab{{\em b\/}}})]{1993Mynicka}
{\sc \au{Mynick, H.~E.}} \yr{1993{\natexlab{{\em b\/}}}}  \at{Transport of energetic ions by low-n magnetic perturbations}.  \jt{Physics of Fluids B: Plasma Physics}  \bvol{5}~(5),  \pg{1471--1481}.

\bibitem[Mynick {\em et~al.\/}(2002)Mynick, Pomphrey \& Ethier]{2002Mynick}
{\sc \au{Mynick, H.~E.}, \au{Pomphrey, N.} \& \au{Ethier, S.}} \yr{2002}  \at{Exploration of stellarator configuration space with global search methods}.  \jt{Physics of Plasmas}  \bvol{9}~(3),  \pg{869--876}.

\bibitem[Najmabadi {\em et~al.\/}(2008)Najmabadi, Raffray, Abdel-Khalik, Bromberg, Crosatti, El-Guebaly, Garabedian, Grossman, Henderson, Ibrahim {\em et~al.\/}]{2008Najmabadi}
{\sc \au{Najmabadi, F.}, \au{Raffray, A.~R.}, \au{Abdel-Khalik, S.~I.}, \au{Bromberg, L.}, \au{Crosatti, L.}, \au{El-Guebaly, L.}, \au{Garabedian, P.~R.}, \au{Grossman, A.~A.}, \au{Henderson, D.}, \au{Ibrahim, A.} \& \au{others}} \yr{2008}  \at{The {ARIES-CS} compact stellarator fusion power plant}.  \jt{Fusion Science and Technology}  \bvol{54}~(3),  \pg{655--672}.

\bibitem[Nazikian {\em et~al.\/}(1997)Nazikian, Fu, Batha, Bell, Bell, Budny, Bush, Chang, Chen, Cheng {\em et~al.\/}]{1997Nazikian}
{\sc \au{Nazikian, R.}, \au{Fu, G.~Y.}, \au{Batha, S.~H.}, \au{Bell, M.~G.}, \au{Bell, R.~E.}, \au{Budny, R.~V.}, \au{Bush, C.~E.}, \au{Chang, Z.}, \au{Chen, Y.}, \au{Cheng, C.~Z.} \& \au{others}} \yr{1997}  \at{Alpha-particle-driven toroidal {Alfv{\'e}n} eigenmodes in the tokamak fusion test reactor}.  \jt{Physical Review Letters}  \bvol{78}~(15),  \pg{2976}.

\bibitem[Nishimura {\em et~al.\/}(2013)Nishimura, Todo, Spong, Suzuki \& Nakajima]{2013Nishimura}
{\sc \au{Nishimura, S.}, \au{Todo, Y.}, \au{Spong, D.~A.}, \au{Suzuki, Y.} \& \au{Nakajima, N.}} \yr{2013}  \at{Simulation study of {Alfv{\'e}n}-eigenmode-induced energetic ion transport in {LHD}}.  \jt{Plasma and Fusion Research}  \bvol{8},  \pg{2403090--2403090}.

\bibitem[Paul {\em et~al.\/}(2022)Paul, Bhattacharjee, Landreman, Alex, Velasco \& Nies]{2022Paul}
{\sc \au{Paul, E.~J.}, \au{Bhattacharjee, A.}, \au{Landreman, M.}, \au{Alex, D.}, \au{Velasco, J.~L.} \& \au{Nies, R.}} \yr{2022}  \at{Energetic particle loss mechanisms in reactor-scale equilibria close to quasisymmetry}.  \jt{Nuclear Fusion}  \bvol{62}~(12),  \pg{126054}.

\bibitem[Rahbarnia {\em et~al.\/}(2020)Rahbarnia, Thomsen, Schilling, Vaz~Mendes, Endler, Kleiber, K{\"o}nies, Borchardt, Slaby, Bluhm {\em et~al.\/}]{2020Rahbarnia}
{\sc \au{Rahbarnia, K.}, \au{Thomsen, H.}, \au{Schilling, J.}, \au{Vaz~Mendes, S.}, \au{Endler, M.}, \au{Kleiber, R.}, \au{K{\"o}nies, A.}, \au{Borchardt, M.}, \au{Slaby, C.}, \au{Bluhm, T.} \& \au{others}} \yr{2020}  \at{Alfv{\'e}nic fluctuations measured by in-vessel {Mirnov} coils at the {Wendelstein 7-X} stellarator}.  \jt{Plasma Physics and Controlled Fusion}  \bvol{63}~(1),  \pg{015005}.

\bibitem[Sigmar {\em et~al.\/}(1992)Sigmar, Hsu, White \& Cheng]{1992Sigmar}
{\sc \au{Sigmar, D.~J.}, \au{Hsu, C.~T.}, \au{White, R.} \& \au{Cheng, C.~Z.}} \yr{1992}  \at{Alpha-particle losses from toroidicity-induced {Alfv{\'e}n eigenmodes. Part II: Monte Carlo simulations and anomalous alpha-loss processes}}.  \jt{Physics of Fluids B: Plasma Physics}  \bvol{4}~(6),  \pg{1506--1516}.

\bibitem[Spong {\em et~al.\/}(2010)Spong, D’azevedo \& Todo]{2010Spong}
{\sc \au{Spong, D.~A.}, \au{D’azevedo, E.} \& \au{Todo, Y.}} \yr{2010}  \at{{Clustered frequency analysis of shear Alfv{\'e}n modes in stellarators}}.  \jt{Physics of Plasmas}  \bvol{17}~(2),  \pg{022106}.

\bibitem[Spong {\em et~al.\/}(2017)Spong, Holod, Todo \& Osakabe]{2017Spong}
{\sc \au{Spong, D.~A.}, \au{Holod, I.}, \au{Todo, Y.} \& \au{Osakabe, M.}} \yr{2017}  \at{Global linear gyrokinetic simulation of energetic particle-driven instabilities in the {LHD} stellaratora}.  \jt{Nuclear Fusion}  \bvol{57}~(8),  \pg{086018}.

\bibitem[Takechi {\em et~al.\/}(2002)Takechi, Matsunaga, Toi, Isobe, Minami, Tanaka, Nishimura, Takahashi, Okamura \& Matsuoka]{2002Kakechi}
{\sc \au{Takechi, M.}, \au{Matsunaga, G.}, \au{Toi, K.}, \au{Isobe, M.}, \au{Minami, T.}, \au{Tanaka, K.}, \au{Nishimura, S.}, \au{Takahashi, C.}, \au{Okamura, S.} \& \au{Matsuoka, K.}} \yr{2002}  \at{{Transition of toroidal Alfven eigenmode to global Alfven eigenmode in CHS heliotron/torsatron plasmas heated by neutral beam injection}}.  \jt{Journal of Plasma and Fusion Research}  \bvol{78}~(12),  \pg{1273--1274}.

\bibitem[Todo {\em et~al.\/}(2017)Todo, Seki, Spong, Wang, Suzuki, Yamamoto, Nakajima \& Osakabe]{2017Todo}
{\sc \au{Todo, Y.}, \au{Seki, R.}, \au{Spong, D.~A.}, \au{Wang, H.}, \au{Suzuki, Y.}, \au{Yamamoto, S.}, \au{Nakajima, N.} \& \au{Osakabe, M.}} \yr{2017}  \at{Comprehensive magnetohydrodynamic hybrid simulations of fast ion driven instabilities in a {Large Helical Device} experiment}.  \jt{Physics of Plasmas}  \bvol{24}~(8),  \pg{081203}.

\bibitem[Toi {\em et~al.\/}(2011)Toi, Ogawa, Isobe, Osakabe, Spong \& Todo]{2011Toi}
{\sc \au{Toi, K.}, \au{Ogawa, K.}, \au{Isobe, M.}, \au{Osakabe, M.}, \au{Spong, D.~A.} \& \au{Todo, Y.}} \yr{2011}  \at{Energetic-ion-driven global instabilities in stellarator/helical plasmas and comparison with tokamak plasmas}.  \jt{Plasma Physics and Controlled Fusion}  \bvol{53}~(2),  \pg{024008}.

\bibitem[Varela {\em et~al.\/}(2021)Varela, Shimizu, Spong, Garcia \& Ghai]{2021Varela}
{\sc \au{Varela, J.}, \au{Shimizu, A.}, \au{Spong, D.~A.}, \au{Garcia, L.} \& \au{Ghai, Y.}} \yr{2021}  \at{{Study of the Alfven eigenmodes stability in CFQS plasma using a Landau closure model}}.  \jt{Nuclear Fusion}  \bvol{61}~(2),  \pg{026023}.

\bibitem[Varela {\em et~al.\/}(2017)Varela, Spong \& Garcia]{2017Varela}
{\sc \au{Varela, J.}, \au{Spong, D.~A.} \& \au{Garcia, L.}} \yr{2017}  \at{Analysis of {Alfv{\'e}n} eigenmode destabilization by energetic particles in {Large Helical Device using a Landau-closure model}}.  \jt{Nuclear Fusion}  \bvol{57}~(4),  \pg{046018}.

\bibitem[Weller {\em et~al.\/}(1994)Weller, Spong, Jaenicke, Lazaros, Penningsfeld \& Sattler]{1994Weller}
{\sc \au{Weller, A.}, \au{Spong, D.~A.}, \au{Jaenicke, R.}, \au{Lazaros, A.}, \au{Penningsfeld, F.~P.} \& \au{Sattler, S.}} \yr{1994}  \at{{Neutral beam driven global Alfv{\'e}n eigenmodes in the Wendelstein W7-AS stellarator}}.  \jt{Physical Review Letters}  \bvol{72}~(8),  \pg{1220}.

\bibitem[White {\em et~al.\/}(2022)White, Bierwage \& Ethier]{2022White}
{\sc \au{White, R.}, \au{Bierwage, A.} \& \au{Ethier, S.}} \yr{2022}  \at{Poor confinement in stellarators at high energy}.  \jt{Physics of Plasmas}  \bvol{29}~(5),  \pg{052511}.

\bibitem[White(2011)]{2011White}
{\sc \au{White, R.~B.}} \yr{2011}  \at{Modification of particle distributions by magnetohydrodynamic instabilities {II}}.  \jt{Plasma Physics and Controlled Fusion}  \bvol{53}~(8),  \pg{085018}.

\bibitem[White(2012)]{2012White}
{\sc \au{White, R.~B.}} \yr{2012}  \at{Modification of particle distributions by {MHD instabilities I}}.  \jt{Communications in Nonlinear Science and Numerical Simulation}  \bvol{17}~(5),  \pg{2200--2214}.

\bibitem[White \& Duarte(2023)]{2023White}
{\sc \au{White, R.~B.} \& \au{Duarte, V.~N.}} \yr{2023}  \at{Assessment of radial transport induced by {Alfv{\'e}nic} resonances in tokamaks and stellarators}.  \jt{Physics of Plasmas}  \bvol{30}~(1),  \pg{012502}.

\bibitem[White {\em et~al.\/}(1983)White, Goldston, McGuire, Boozer, Monticello \& Park]{1983White}
{\sc \au{White, R.~B.}, \au{Goldston, R.~J.}, \au{McGuire, K.}, \au{Boozer, A.~H.}, \au{Monticello, D.~A.} \& \au{Park, W.}} \yr{1983}  \at{Theory of mode-induced beam particle loss in tokamaks}.  \jt{The Physics of Fluids}  \bvol{26}~(10),  \pg{2958--2965}.

\bibitem[Yamamoto {\em et~al.\/}(2007)Yamamoto, Nagasaki, Suzuki, Mizuuchi, Okada, Kobayashi, Blackwell, Kondo, Motojima, Nakajima {\em et~al.\/}]{2007Yamamoto}
{\sc \au{Yamamoto, S.}, \au{Nagasaki, K.}, \au{Suzuki, Y.}, \au{Mizuuchi, T.}, \au{Okada, H.}, \au{Kobayashi, S.}, \au{Blackwell, B.}, \au{Kondo, K.}, \au{Motojima, G.}, \au{Nakajima, N.} \& \au{others}} \yr{2007}  \at{{Observation of magnetohydrodynamic instabilities in Heliotron J plasmas}}.  \jt{Fusion Science and Technology}  \bvol{51}~(1),  \pg{92--96}.

\bibitem[Zonca {\em et~al.\/}(2015)Zonca, Chen, Briguglio, Fogaccia, Vlad \& Wang]{2015Zonca}
{\sc \au{Zonca, F.}, \au{Chen, L.}, \au{Briguglio, S.}, \au{Fogaccia, G.}, \au{Vlad, G.} \& \au{Wang, X.}} \yr{2015}  \at{Nonlinear dynamics of phase space zonal structures and energetic particle physics in fusion plasmas}.  \jt{New Journal of Physics}  \bvol{17}~(1),  \pg{013052}.

\end{thebibliography}

\appendix

\section{Analysis of the resonance condition}
\label{app:resonance_condition}

Under the assumption of a vacuum magnetic field, the perturbed radial equation of motion reads,
\begin{align}
   \delta  \dot{\psi} &= \frac{\bm{B}_0 \times \nabla \delta \Phi \cdot \nabla \psi}{B_0^2} + v_{\|} \frac{\delta \bm{B} \cdot \nabla \psi }{B_0} = 
    -\left(\omega + (\iota m - n) \dot{\zeta} \right) \frac{\delta \Phi_{,\theta}}{\omega},
    \label{eq:radial_drift}
\end{align}
where subscript commas denote partial derivatives. The toroidal motion, $\dot{\zeta} = B_0v_{\|}/G$, is not modified by the perturbation. The first term accounts for the perturbed $\bm{E} \times \bm{B}$ drift, while the second term accounts for streaming along the perturbed magnetic field. 
Given the form for the perturbed potential \eqref{eq:delta_phi}, the perturbed radial drift is proportional to $\cos(\eta)$. We begin our analysis by expressing $\cos(\eta)$ along the unperturbed trajectory in terms of Bessel functions \eqref{eq:bessel_sin_eta}. This expression is used to evaluate the perturbed radial drift in \eqref{eq:net_radial_drift}-\eqref{eq:psi_cs_app} and the resulting phase-space island width \eqref{eq:island_width_app}. 

We begin with the expressions for the unperturbed trajectories \eqref{eq:unperturbed_drifts}. 
Due to the relatively small magnitude of the periodic contributions, $\chi_j$ and $\zeta_j$, in comparison to the secular terms, $\omega_{\chi}$ and $\omega_{\zeta}$, 
the trajectories can be approximated as:
\begin{align}
 \left\{
    \begin{array}{l}
      \displaystyle
     \chi 
     \approx \chi^0 + \omega_{\chi}t + \sum_{j\ne 0} \frac{\chi_j}{j\omega_{\chi}} \sin(j \left[\chi^0 + \omega_{\chi}t\right]), \\
  \displaystyle  \zeta 
  \approx \zeta^0 + \omega_{\zeta}t + \sum_{j\ne 0} \frac{\zeta_j}{j\omega_{\chi}} \sin(j\left[\chi^0 + \omega_{\chi}t\right]),
      \end{array}
    \right.
\end{align}
where $\chi^0 = \chi(t = 0)$ and $\zeta^0 = \zeta(t = 0)$. 

To evaluate the radial drift due to the perturbation \eqref{eq:radial_drift}, we evaluate the phase along the unperturbed trajectory
\begin{align*}
\eta(t) = \eta^0 + \left(m\omega_{\chi} - (n - Nm)\omega_{\zeta} + \omega \right) t + \sum_{j\ne 0}\left[ \frac{m\chi_j}{j\omega_{\chi}}- (n-Nm) \frac{\zeta_j}{j\omega_{\chi}}\right]\sin(j \left[\chi^0 +\omega_{\chi}t\right]),
\end{align*}
where $\eta^0 = m \theta^0 - n \zeta^0$ is the initial phase. 
Under the assumption that one term, $j = j'$, is dominant in the expression for the drifts \eqref{eq:unperturbed_drifts}, the phase factor $\cos(\eta)$ can be expressed in terms of Bessel functions of the first kind, 
\begin{align}
    \cos(\eta) &= \sum_{k} J_k\left( \eta_{j'}\right) \cos \left(\eta^0 + \Omega_{kj'}t \right),
    \label{eq:bessel_sin_eta}
\end{align}
where $\eta_{j'} = \left(m\chi_{j'}- (n-Nm)\zeta_{j'}\right)/(j'\omega_{\chi})$ and $\Omega_{kj'} = (m+kj')\omega_{\chi} - (n - Nm)\omega_{\zeta} + \omega$. 

By expressing the instantaneous perturbed radial drift \eqref{eq:radial_drift} in terms of $\cos(\eta)$ along the unperturbed trajectory \eqref{eq:bessel_sin_eta} we obtain the expression:
\begin{multline}
    \delta \dot{\psi} = \sum_{k} \Bigg(\psi_{kj'}^{0} \cos\left(\Omega_{kj'} t + \eta^0 \right) + \psi_{kj'}^{+} \cos \left(\Omega_{kj'} t + \eta^0 - j' \chi^0\right) \\
    + \psi_{kj'}^{-} \cos \left(\Omega_{kj'} t + \eta^0 + j' \chi^0\right) \Bigg)
    \label{eq:net_radial_drift}
\end{multline}
with amplitudes given by,
\begin{align}
\left \{ \begin{array}{l}
\psi_{kj'}^{0} = - m \hat{\Phi} J_k(\eta_{j'}) \left(1 + (\iota m-n) \frac{\omega_{\zeta}}{\omega} \right), \\ \displaystyle
\psi_{kj'}^{\pm} = -  \frac{m \hat{\Phi} J_{k\pm 1}(\eta_{j'})}{2} (\iota m-n) \frac{\zeta_{j'}}{\omega}  . 
\end{array} \right. 
\label{eq:psi_cs_app}
\end{align}
For a net drift to occur, a resonance condition $\Omega_{kj'} = 0$ must be satisfied. The Fourier amplitudes $\psi_{kj'}^0$ and $\psi_{kj'}^{\pm}$ then determine the amplitude of the radial transport. The associated full island width is given by,
\begin{align}
    w^{\psi}_{kj'} = 2\sqrt{\left | \frac{\psi_{kj'}}{\Omega_{kj'}'(\psi)} \right |} \approx 2\sqrt{\left | \frac{\psi_{kj'}}{(m + kj')\omega_{\theta}'(\psi)} \right |},
    \label{eq:island_width_app}
\end{align}
where $\psi_{kj'}$ is taken to be $\psi_{kj'}^0$ or $\psi_{kj'}^{\pm}$. 

We now consider the scaling of the cosine amplitudes \eqref{eq:psi_cs_app} with respect to the radial perturbed magnetic field amplitude and magnetic drifts. If a resonance condition $\Omega_{kj'} = 0$ is satisfied, the corresponding radial perturbed field scales with the perturbation amplitude and mode numbers as $\delta \hat{B}^{\psi} \sim m (\iota m - n) \hat{\Phi}$. Using the resonance condition, the factor appearing in $\psi_{kj'}^0$ can be expressed as $1 + (\iota m -n )\omega_{\zeta}/\omega = (-kj' \omega_{\chi} + m (\iota \omega_{\zeta}-\omega_{\theta}))/\omega$. Since the quantity $\iota - \omega_{\theta}/\omega_{\zeta}$ scales with the magnitude of the magnetic drifts, at low energy, this factor approaches $kj' \omega_{\chi}/\omega$. Thus, $\psi_0^0 \sim (\iota - \omega_{\theta}/\omega_{\zeta}) J_0(\eta_{j'}) \delta \hat{B}^{\psi}$. The scaling of each harmonic is summarized in Table \ref{tab:harmonic_scaling}. Considering the small argument limit of $\psi_0^{\pm}$, this harmonic will scale with two factors of the magnetic drifts through both $\eta_{j'}$ and $\zeta_{j'}$. Thus $\psi_0^0$ will dominate over $\psi_0^{\pm}$ for $k = 0$.
Considering $k \ne 0$, $\psi_{k}^0$ will have the same scaling with the magnetic drifts as $\psi_k^{-}$ for $k > 0$, while $\psi_k^{+}$ will be smaller in magnitude by a factor of the magnetic drifts. The magnitude of the harmonics decreases strongly with $|k|$ due to the Bessel coupling parameter. When considering the dependence on the magnetic drifts, the most significant radial transport will arise from $\psi_0^0$, $\psi_{\pm j'}^0$, and $\psi_{\mp j'}^{\pm }$. Considering the limit of large mode number perturbations, $\psi_{\pm j'}^0$ will dominate due to its dependence on $\eta_{j'}$.

\begin{table}
    \centering
    \begin{tabular}{|c|c|c|}
       $\psi_0^0$  & $\psi_{k\ne 0}^0$ & $\psi_{k}^{ \pm}$ \\ \hline 
        $(\iota - \omega_{\theta}/\omega_{\zeta}) J_0(\eta_{j'}) \delta \hat{B}^{\psi}$ & $ J_k(\eta_{j'})\delta \hat{B}^{\psi}$ & $ J_{k\pm 1}(\eta_{j'}) \zeta_{j'} \delta \hat{B}^{\psi}$
    \end{tabular}
    \caption{Scaling of cosine harmonic amplitudes of the perturbed radial drift \eqref{eq:psi_cs_app} with the radial perturbed magnetic field, magnetic drifts, and mode numbers. The quantities $(\iota - \omega_{\theta}/\omega_{\zeta})$ and $\zeta_{j'}$ scale with the drift amplitude. The quantity $\eta_{j'}$ scales with both the drift amplitude and the mode numbers. }
    \label{tab:harmonic_scaling}
\end{table}

\end{document}